\documentclass[useAMS,usenatbib,usegraphicx]{mn2e}
\def\la{\raise.5ex\hbox{$<$}\kern-.8em\lower 1mm\hbox{$\sim$}}
\def\ma{\raise.5ex\hbox{$>$}\kern-.8em\lower 1mm\hbox{$\sim$}}

\def\kms{$\rm km\, s^{-1}$}
\def\cm3{$\rm cm^{-3}$}

\def\Vs{$\rm V_{s}$}
\def\n0{$\rm n_{0}$}
\def\B0{$\rm B_{0}$}

\def\Fh{$\rm F_{h}$~}

\def\mum{$\mu$m}

\def\L12{L$_{12\mu m}$~}
\def\F12{F$_{12\mu m}$~}
\def\agr{a$_{gr}$}

\title[The Infrared Continuum of Active Galaxies]{The Infrared Continuum
of Active Galaxies}
\author[M. Contini, S.M. Viegas, \& M.A. Prieto]{ M. Contini$^{1}$, S.
M. Viegas$^{2}$,and M. A. Prieto$^{3}$\\
$^{1}$School of Physics and Astronomy, Tel Aviv University, Tel Aviv
69978, Israel \\
$^{2}$Instituto de Astronomia, Geof\'{\i}sica e Ci\^encias
Atmosf\'ericas - USP, Brazil\\
$^{3}$Max-Planck Institut f\"{u}r Astronomie, K\"{o}nigsthul 17,
D-69117, Heidelberg, Germany \\}

\begin{document}

\date{Accepted: Received ; in original form 2003 month day}

\pagerange{\pageref{firstpage}--\pageref{lastpage}} \pubyear{2003}

\maketitle

\label{firstpage}

\begin{abstract}

We discuss the different physical processes contributing to the
infrared continuum of AGN, assuming that both photoionization from the
active center and shocks ionize and heat the gas and dust contained in
an ensemble of clouds surrounding the nucleus.
In our model, radiation transfer of primary and secondary radiation
throughout a cloud is calculated consistently with collisional processes
due to the shock. We consider that
the observed continuum corresponds
to reprocessed radiation  from both dust and gas  in the clouds.
Collisional processes are important in the presence of shocks.  The
grains are sputtered crossing the shock front.  The models are
constrained by sputtering as well as by the far-infrared data. The
model is applied to the continuum of Seyfert galaxies from which best
estimate of the nuclear, stellar subtracted, emission is available.
The results show that radiation-dominated
high velocity clouds are more numerous
in Seyfert 1-1.5 whereas shock-dominated low velocity clouds are
dominant in Seyfert type 2. This result is in full agreement with
the unified model for AGN, by which high velocity clouds, placed
deeper into the central region and therefore
reached by a more intense radiation, should play a more significant
role in the spectra of broad line objects. We could therefore conclude
that in type 2 objects, radiation is partly suppressed by
a central dusty medium with a high dust-to-gas ratio.
Once the model approach is tested, a grid of models is used to provide
a phenomenological analysis of the observed infrared spectral energy
distribution. This empirical method is a useful tool to rapidly access
the physical conditions of the AGN emitting clouds. For this,
analytical forms are derived for the two processes contributing to the
infrared emission: dust emission and thermal bremsstrahlung produced
by the NLR clouds. Their relative contribution provides a measurement
of the dust-to-gas ratio.

\end{abstract}


\section{Introduction}

The presence of a dusty torus around the central engine was proposed by
Antonucci \& Miller (1985)  to explain the broad emission-lines
observed in the polarized light of Seyfert 2 galaxies. The torus absorbs
a large fraction of the nuclear emission and reradiates it in the infrared.
This is the basis of the so-called AGN unification scheme largely
analysed  in the literature and that provides a geometrical explanation
for the observed differences between broad line and narrow line objects.

After the pioneering work by Edelson  \& Malkan (1986),
the characteristics of the continuum due
to dust heated by the optical-UV nuclear radiation
has been analyzed by several authors (Barvainis 1992, Pier \& Krolik 1992;  
Laor \& Draine 1993; Granato \& Danese 1994;
Granato, Danese \& Franceschini 1997,  Nenkova, Ivezi\'c \& Elitzur 2002).

The infrared emission  is generally associated to a dusty torus
surrounding the central source of the AGN. Although proposed by
Pier and Krolik (1992,1993), the calculation of the radiative transfer
radiation throughout a clumpy   torus has only been  recently performed by
Nenkova et al. (2002). The AGN obscuring region is modelled
as a toroidal distribution  of dusty clouds placed at
different radii from the
center, each with  same optical depth.
The optically thick clumps are heated by  the nuclear radiation
as well as from the radiation originated in other clumps. The dust
temperature is much higher in their illuminated side.
The authors claim  that heating by diffuse
radiation is highly inefficient and can be neglected.
Following their results, a clumpy torus accounts for   both,
the often observed broad IR bump extending up to 100 \mum ~and  the absence
of the 10 \mum ~silicate feature in type 1 - on axis view- objects.

In general,  two type of models are used in the literature:
a compact torus (radial dimension
smaller than a few pc) with a large optical depth, and  a more extended
(up to hundreds pc) moderately thick disk.
A common characteristic  to all  is that dust survival
in the vicinity of the AGN is set by its evaporation limit. Therefore,
dust can reach temperatures up to 1000-1500 K.
Overall, these models provide
a reasonable fit to the observed spectral energy distribution (SED)
The torus
models usually assume a high equatorial opacity leading to  flat
spectra for type 1 and steep spectra for type 2 galaxies. However,
as pointed out by Alonso-Herrero et al. (2003),
the large number of spectral indices observed in both types
of galaxies, as well as the large number of intermediate
and type 1 objects  with  steep SEDs, can not be explained by
the torus model.

Other simpler descriptions of the IR SED include e.g., that  by
Blain, Barnard, \& Chapman (2003)  who describe
the far-IR to submillimeter  SED of
dusty galaxies  with different forms of  modified black-body models.

Although most available models provide in a way or other a fair
representation of the IR SED, we believe that a more sounded physical
approach is needed.
An important problem affecting all  these models
 - those mentioned  above, as well as 
 in the case of Blain's et al. approach,
 is that  the results
are independent of the dust-to gas ratio. 

For several years, we are pursuing the development of
self-consistent models aimed at explaining both the emission-line and
SED  of AGN.
These models have proved to be successful
in a number of representative  Seyfert galaxies (Contini, Prieto,
Viegas 1998a,b, Contini, Viegas, \& Prieto 2002).
In our modelling approach, the infrared-optical continuum is due to dust
emission and free-free emission from the clouds powered  by the
nuclear radiation, diffuse radiation  and by shocks.
Shocks are due to the cloud motion through a
dilute gas (Contini \& Viegas-Aldrovandi 1990, Viegas \& Contini 1994,
Contini \& Viegas 2000). An important aspect considered 
in that approach  are the conditions
of dust survival when effects as sputtering and
grain-gas collisions are taken on board.
Thus, these models offer an alternative to the
torus model, particularly for the objects showing
infrared emission as extended as the narrow-line region.

Overall,  we consider a model to be  well  constrained 
when both the complete SED and the  emission-line
spectrum are simultaneously modeled. When possible, that
has been our approach, e.g. in the references quoted above. However,
in this paper, we  focus on the optical-IR continuum  only, since 
in this wavelength range the   best estimate
of the nuclear flux emission  and of shock velocities for a reasonable number
of Seyfert galaxies of different types (the sample by
Alonso-Herrero et al. 2003) is available. 

The  paper is structured as follows.
A  short review of the processes involving dust is given in Sect. 2.
Numerical simulations accounting for the presence of both shocks and
photoionization in the NLR are presented  and applied to
the  SEDs of Seyfert  galaxies, described in
Alonso-Herrero et al. (2003),  in Sect. 3.
Concluding remarks appear in Sect. 4.
An empirical method  derived from the models,  aimed at  providing
a first assessment of the
nature of the observed IR  continuum, is  presented in the Appendix.

\section{Dust and Gas emission}

When the physical conditions of a dust cloud is due to a combination
of photoionization and shock, the dust heating and temperature will
depend on the UV radiations as well as on the post-shock temperature.

The  maximum grain temperature, which would account for the
near infrared emission, is limited by sublimation. An estimate
for some Seyfert 2 galaxies shows that the critical distance for
grain survival, i.e., the distance where the grain temperature is
equal to the evaporation temperature, for both graphite 
and silicate, is in the range 0.25 to 50 pc, 
depending on the luminosity of the central source
 (Vaceli et al. 1993).
In addition,
the value of the temperature reached by the grains
depends on the presence of small grains and
on a strong UV radiation from the central source.

Several lines of evidence are  pointing towards the destruction of
very small grains in regions of high UV radiation intensity
(Rowan-Robinson  1992). Destruction of dust by
those effects are relevant when gas velocities  about or
larger than 300 \kms are present; still, velocities of
$\geq$ 100 \kms can have an effect on dust destruction as we shall see
below.

In the following, the effect of photoionization and shocks
on the temperature and survival of dust grains in an emitting
cloud of the NLR  is reviewed.

\subsection{Description of our models: the code SUMA}

Our models result from numerical simulations with
the code SUMA  (Viegas \& Contini 1994,  Viegas \&  Contini 1997,
and references therein)
which requires the following input parameters:
the shock velocity, \Vs, the preshock density, \n0,
the preshock magnetic field, \B0, an ionizing radiation spectrum,
the chemical composition of the gas and a dust-to-gas ratio $d/g$.
As ionizing continuum,  a power-law  characterized by the ionizing
flux at the Lyman limit \Fh
(in photons cm$^{-2}$, s$^{-1}$, eV$^{-1}$) and a spectral index, or
a combination of spectral indices in different wavelength ranges, is
considered.
A plan-parallel symmetry is adopted with the calculations starting
at the shocked edge  and the cloud divided in slabs.
The number of slabs is chosen according to the precision required by
the calculation of the physical conditions.
The geometrical thickness of the cloud, D, is also
an input parameter in the case where shock and photoionization
act on the opposite edges of the cloud. This is the common case
for clouds flowing outward from the nucleus.

The models discussed here have been obtained assuming a spectral
index of 1.5, cosmic chemical abundances for  He, C, N, O, Ne,
Mg, Si, S, Ar, and Fe  (Allen 1973), and \B0 = 10$^{-4}$ gauss.  The other
input parameters are variable in ranges established by the observations
as well as by previous modelling of the NLR.

As the calculations account for the sputtering in the different zone
downstream of the shock front (Viegas \& Contini 1994), the distribution of
the grain sizes along the cloud is automatically derived by SUMA
starting from an initial size. In AGN, carbon and PAH grain features
are relatively less prominent than, for example, in  starburst galaxies,
therefore only silicate grains are considered (Siebenmorgen et al
2003).

Self-consistent calculations lead to the continuum and emission-line
spectra which are compared with  the data.  

\subsection{Description of our models : the IR emission}

\begin{figure}
\includegraphics[width=78mm]{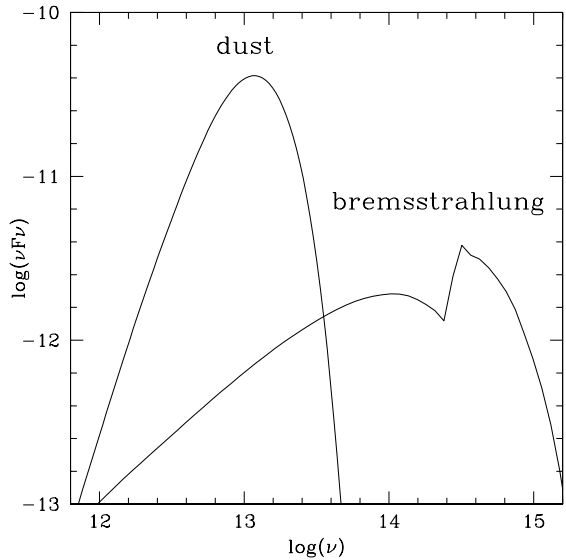}
\caption{The dust reemission peak relative to bremsstrahlung emission
for a prototype model M3 (Table 1)}
\end{figure}

As Nenkova et al. we also consider an
ensemble of dusty clouds placed at different distances from the active
center. In our model, radiation transfer of primary and secondary radiation
throughout a cloud is calculated consistently with collisional processes
due to the shock. Moreover, we consider that
{\it the observed continuum corresponds
to reprocessed radiation  from both dust and gas  in the clouds.}
For NLR clouds moving outwards, in the edge of the cloud facing the
photoionizing source, dust and gas are mainly heated by the primary
radiation, although the effect of the diffuse radiation from the
ionized gas cannot be neglected.  Collision effects are less important
because they increase with temperature, which is usually smaller in
the photoionized region.  The temperature of the gas, in fact, does
not exceed 2-3 10$^4$ K in this region.

On the other hand, at the opposite edge, the gas is collisionally
heated to relatively high temperatures by the shock.  The post-shock
temperatures depend on the shock velocity ($\propto$ \Vs$^2$).  In
these models, dust and gas are coupled throughout the shock front.
Collision phenomena dominate at high temperature leading to relatively
high temperatures of gas and grains.  Diffuse radiation from the hot
gas bridges between the shock dominated and the radiation dominated
sides of the cloud.

The structure of the gas downstream is determined by the shock.  In
fact, the density distribution results from compression and the
temperature distribution is strongly affected by the cooling rate.
Due to gas and dust mutually heating and cooling, the maximum
temperature of dust depends on the shock velocity, which therefore
determines the frequency of the reradiation peak in the infrared.  An
analytical relation between the wavelength of the maximum intensity of
dust reradiation in the IR spectrum and shock velocity is given by
Draine (1981).

The  hot NLR clouds  produce  bremsstrahlung. Depending
on the temperature distribution in the cloud, this mechanism may
contribute to the infrared continuum
(see, for instance, Contini \& Viegas-Aldrovandi 1990, fig 2c).
{\it The intensity of dust IR emission
relative to bremsstrahlung depends on the  dust-to-gas  ratio
d/g.} Usually the dust contribution dominates in the far infrared
range but in the near-IR, both processes may contribute to the
continuum  as shown in Fig. 1, where the single-cloud
continuum of model M3 (Table 1) is plotted.

The behaviour at high frequency of
the free-free emission is largely dependent on the  gas velocities.
In Fig. 1,
it shows a linear behavior at low frequencies, then  a local maximum,
then a maximum at a higher frequency. However, for shock-dominated
clouds, the
bremsstrahlung  shows only a
maximum at higher frequencies.

Those are  the most common types of continuum
obtained from a single-cloud model.
Very seldom the
line and continuum spectra of the galaxies can be reproduced by
single-cloud models; on the other hand, multi-cloud models are
generally required to explain the multiwavelength spectrum
and are obtained as a
weighted average of the single-cloud models. The characteristics of
some of the most common single-clouds found in our previous
analysis are listed in Table 1. Multi-cloud models lead  to IR bumps
wider than black body curves.  Actually, in some compact objects,
e.g. Mrk 3, Mrk 34, and Mrk 78 (Contini \& Viegas 2000) dust emission
is associated to a single velocity model.  This is in contrast with
the modelling of Circinus SED (Contini et al.), which requires clouds with
different velocities contributing at  different frequencies in the
infrared  range. That suggests that images of Circinus in the
different IR wavelengths may show very different.

Previous modelling of the representative galaxies with SUMA (e.g. NGC
5252, Circinus, NGC 4151) provided us with hints on
which types of clouds are predominant in the NLR (Contini,
Viegas \& Prieto 1998a,b, Contini \& Viegas 2000).  Adopting an
initial   grain size of 0.20 \mum, it was found that \Vs~ ranges
between 100 and 1000 \kms, \n0 between 100 and 1000 \cm3 , log(\Fh)
between 8 and 13, and D between 0.01 and 10 pc.

\begin{figure}
\includegraphics[width=78mm]{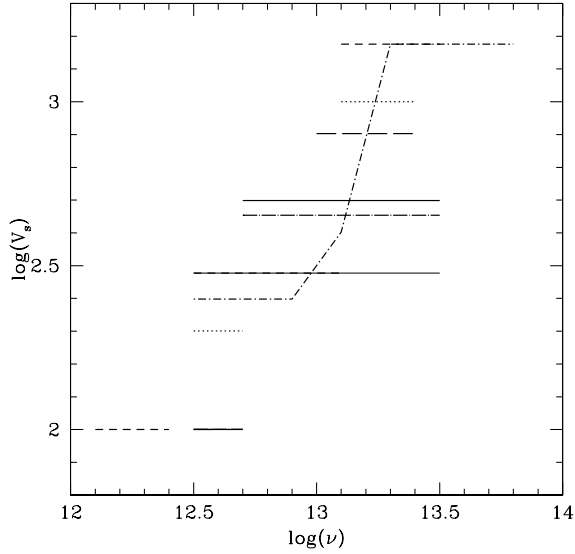}
\caption
{
\Vs  ~adopted  to model  the IR data.
NGC 7130 (dotted line), NGC 4151 (short dash),
NGC 4051 (long dash), Circinus (short dash-dot),
NGC 5252 (long dash-dot), NGC 4388 (solid).
Mrk 3, Mrk 34, Mrk 78 (thin solid) are selected
from  (Contini \& Viegas 2000).
}
\end{figure}

\begin{figure}
\includegraphics[width=78mm]{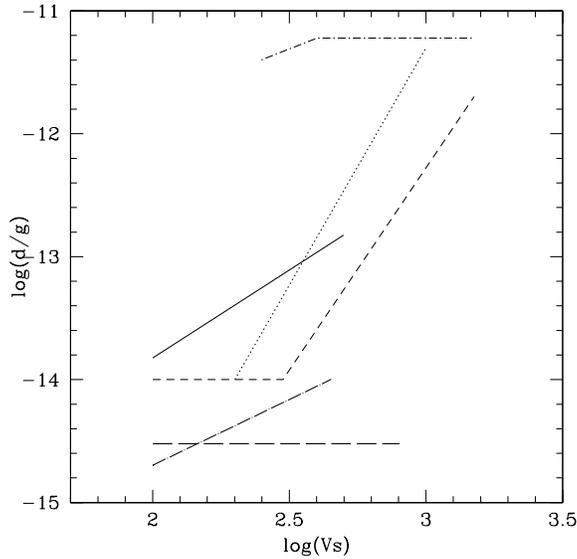}
\caption
{d/g ratios as a function of \Vs ~adopted to model single galaxies.
Symbols as in Fig. 2
}
\end{figure}

The IR bump results from summing up the SED contribution
of dusty clouds with different velocities within the NLR,
each of them peaking at the corresponding frequency.
The maximum of dust emission  occurs in the mid-IR for
relatively high velocity clouds. The dust-to-gas
ratio may vary from cloud to cloud. From the results obtained
for Seyfert galaxies previously modelled by SUMA, we show in
Fig. 2, the frequency range corresponding to dust emission
associated to a given shock velocity. A similar plot is shown in
Fig. 3 for  d/g. There is a tendency to find higher d/g for
higher shock velocities.

\subsection{Sputtering of dust grains}

\begin{figure}
\includegraphics[width=78mm]{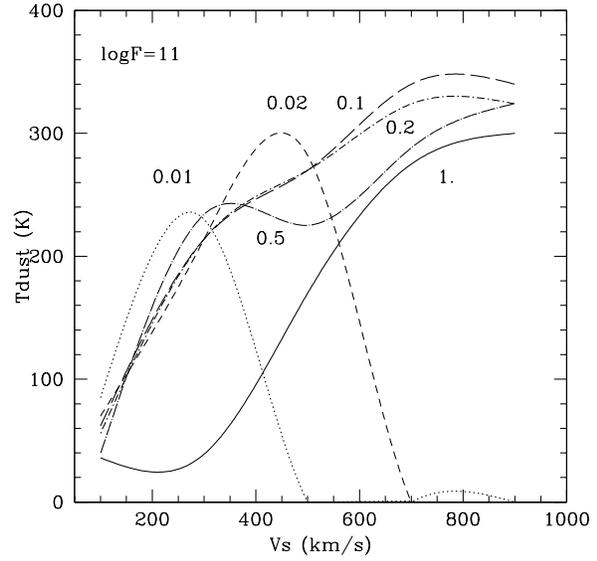}
\includegraphics[width=78mm]{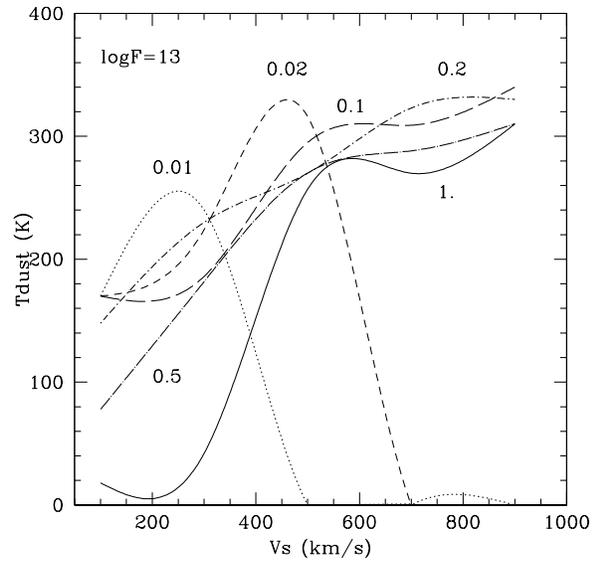}
\caption
{Effective temperature of dust as a function of the shock velocity for
different a$_{gr}$ (in \mum). Dotted lines : a$_{gr}$=0.01 \mum,
short-dashed : 0.02 \mum, long-dashed : 0.1 \mum, short-dash-dot : 0.2 \mum,
long-dash-dot : 0.5 \mum, solid : 1 \mum.
Top panel : log \Fh=11, bottom panel : log \Fh=13.
}
\end{figure}

In models accounting for the presence of  shocks, sputtering
changes the initial grain-size distribution, creating
a deficiency of small grains with radius a$_{gr}$ $<$ 50 \AA  ~compared
to their preshock
abundances (Dwek, Foster, \& Vankura 1996).
The altered grain size distribution depends on shock velocity
and on the density of the medium.
The sputtering rate increases when the dust is in motion.
At temperatures below 10$^5$ K, all the sputtering is caused
by the relative gas-grain motion.

When sputtering is present,
the lifetime of grain survival depends on the
distance downstream covered by the grains.
In order to calculate the  critical distance,
we use the formula from Shull (1978) which gives the rate of change
of grain velocity behind the shock front:

\noindent
dv$_{gr}$/dt=(v$_{gr}$/2B)(dB/dt)-($\pi a_{gr}^2$/m$_{gr}$)($\rho
v_{gr}^2$)-F$_{pl}$/m$_{gr}$

\noindent
Immediately behind the shock front the grain's
gyrovelocity about the magnetic field B is v$_{gr} \sim$ 3/4 \Vs.

The first term on the right hand side is the rate of change of the grain's
gyrovelocity calculated from the conservation of the adiabatic invariant
v$_{gr}^2$/B (Dwek 1981). For a  "frozen in" magnetic field
(see Contini \& Shaviv 1982):

\noindent
(v$_{gr}$/2B)(dB/dt) = (v$_{gr}$/2n)(dn/dt),  (B=\B0 n/\n0)

\noindent
The second term contains the collisional drag force resisting the
grain motion
(Dwek 1981) and the third term  shows the plasma drag F$_{pl}$ (Shull
1978). Neglecting the plasma drag, if
(v$_{gr}$/2B)(dB/dt) $<$ ($\pi$ a$_{gr}^2$/m$_{gr}$)($\rho$ v$_{gr}^2$)

\noindent
the grains decelerate and sputtering is prevented.

Considering that  ds=dt v$_{gas}$ is the differential distance
corresponding
to lifetime for evolution of the gas throughout  a  slab,  and
adopting
a  silicate density  $\sim$ 3 gr \cm3,  we obtain :
8/3 (dn/n$^2$) a$_{gr}$/m$_{gas}$ $<$ ds,

\noindent
and integrating, the critical distance is given by:

s (cm) $>$  10$^{16}$  a$_{gr}$ (\mum) / n (10$^4$ \cm3)

To perform the calculations a cloud is divided into a
number of slabs (up to 300) where the
physical conditions are  considered  homogeneous.
The geometrical thickness of
the slabs is automatically calculated by SUMA
according to the gradient of the temperature downstream, in order to
calculate
as smoothly as possible
the distribution of the physical conditions throughout the
cloud.  The first slabs in the
immediate postshock region show the maximum temperature which
depends on \Vs. These slabs can be relatively large because
recombination coefficients are lower the higher the temperature.
The geometrical thickness  of the slab closest to the shock front,
$\delta$, is therefore determined by the model.

Comparing $\delta$ with the critical distance for
models with low \Vs ($\sim$ 100 \kms), a preshock density \n0=100-300
\cm3 (corresponding to a density n $\simeq$ 1000 \cm3 downstream),
grains with \agr=10$^{-5}$ cm are decelerated and hardly sputtered
if $\delta  > $ 10$^{16}$ cm.  For higher n ($>$ 10$^4$ \cm3) and a stronger
shock (\Vs$>$300 \kms), the critical distance is s $\simeq$ 10$^{15}$
cm.  Actually, the models show slab widths $<$ 10$^{15}$ cm for \Vs
$\leq$ 700 \kms. Since the physical conditions in the slab are assumed
to be constant, sputtering may be efficient.

For high velocities (\Vs $\sim$ 1000 \kms) temperatures are
high. Recall that da$_{gr}$/dt $\propto$ (T/10$^6$)$^{2.8}$ (Draine \&
Salpeter 1979), so small grains can be easily destroyed.  Therefore, we
adopt
large grains and very small slabs ($\delta  <$ 10$^{13}$ cm) for grain
survival. Narrow slabs improve the precision of the calculations.
Such slab widths are also much smaller than the critical distances s
which are $\sim$ 10$^{16}$ cm and $\sim$ 10$^{14}$ cm for \n0=300 \cm3
and \n0=1000 \cm3, respectively.

The effective temperature of dust calculated for different
shock velocities and different
grain sizes is presented in Fig. 4. Notice that for
\Vs $\geq$ 700 \kms only grains with
\agr $>$ 0.1 \mum ~survive sputtering and contribute to heating the
grains to
relatively high temperatures.

\section{The near- and mid- infrared continuum of a sample of galaxies}

\begin{table}
\caption{The input parameters of single-cloud models}
\begin{tabular}{lllllllll}\\ \hline
\ model & M1 & M2 & M3 & M4 & M5 & M6 & M7 \\ \hline \\
\Vs (\kms) & 100 & 500 & 500 & 500 & 500 & 700  & 1000 \\
\n0 (\cm3) & 1000 & 300 & 300 & 300 & 300 & 300 & 1000 \\
log(\Fh)    & - & 11 & 11& 11& - & - & 11 \\
D (10$^{19}$ cm)& 1 & 1& 1& 1& 1 & 1& 0.01 \\
a$_{gr}$ (\mum) &  0.2 & 0.2 & 0.2 & 0.2 &0.2& 0.5 & 1.\\
d/g (10$^{-14}$)$^1$ & 1 & 1 & 10 & 100 & 10 & 800 & 10 \\
\hline\\
\end{tabular}

$^1$ by number, corresponding to 4 10$^{-4}$ by mass for silicates

\end{table}

\begin{table*}  
\caption{The models in Figs. 5 and 6}  
\begin{tabular}{lllllllll}\\ \hline  

Galaxy &  Seyfert type   &  models$^a$ & relative weights$^b$ & CF$^c$
\\ \hline  
Mkn 334 & 1.8 &  M1, M2, M3 &  0, -4.1 , -4.6   & -8.9 \\  
Mkn 573 & 2. &  M1 &  0 & -8.9\\  
UM 146 & 1.9 &  M1, M2, M3, M4 &  0, -3.8 , -4.6 , -4.1 & -10.\\  
NGC 1068 & 2.-1.8 &  M1, M5 & 0  , -2.7 &  -6.9\\  
NGC 3362 & 2. &  M1, M4 & 0  , -4. &  -10.5 \\  
NGC 3786 & 1.8 & M1, M3, M7 &  0 , -4.5 , -4.3 & -9.5\\  
NGC 4579 & 1.9 &  M1, M2 &  0 , -3.75 &  -9.1\\  
NGC 5033 & 1.9-1.5 &  M1, M3,  M7 &  0 , -4.5 ,  -4.6 &  -9.4\\  
NGC 5194 & 2. &  M1, M3 &  0 , -5.1 &  -9.5\\  
NGC 5273 & 1.9-1.5 &  M1, M3 &  0 , -4.13 &  -9.7\\  
NGC 5347 & 2. &  M1 &  0 &  -9.0\\  
Mkn 471 & 1.8 &  M1, M7 &  0 , -4.8   &  -10.\\  
NGC 5674 & 1.9 &  M1, M3, M7 &  0 , -4. , -5.3   & -9.5 \\  
UGC 12138 & 1.8 &  M1, M3, M7 &  0 , -4.4 , -2.1 &  -9.4 \\  
NGC 7479 & 1.9 &  M1, M6 &  0 , -3.3   &  -9.6 \\  
NGC 7674 & 2. &  M1, M3, M5 &  0 , -5.3 , -3.   &  -8.5 \\  
&  & \\  
Mkn 335 &  1.0 &  M1, M2, M3 &  0 , -3.6 , -3.6 &  -9.1 \\  
Mkn 590 &  1.0 &  M1, M3 &  0 , -4.6 &  -8.8 \\  
NGC 3227 &  1.5 &  M1, M3 &  0 , -4.2 &  -8.5\\  
Mkn 766 &  1.5 &  M1, M3 &  0 , -3.9   &  -8.7 \\  
NGC 5548 &  1.5 &   M1, M2, M3 &  0 , -4. , -4.   &  -8.7 \\  
Mkn 841 &  1.5 &   M1, M3 &  0 , -4.1   &  -9.1 \\  
NGC 7469 &  1.0 &  M1, M2, M7 &  0 , -3.95 , -4.6 &  -8.4 \\  
NGC 7603/ Mkn 530 &  1.5/1.8 &  M1, M2, M3 &  0 , -3.8 , -4.5 &   -8.9\\  

\hline\\  
\end{tabular}  

\flushleft  
$^a$  The models are represented in Fig. 5 with the following symbols :  
M1 : long-dashed; M2 : solid ; M3 : short-dashed ; M4 :
short-dash-dotted ;  
M5 : long-dash-dotted ; M6 and M7 : dotted.  

$^b$ in logarithm  

$^c$ The conversion factor (in logarithm) for each galaxy which accounts
for the distance of  
the emitting clouds to the active center and for the distance of the  
galaxy to Earth. To compare the multi-cloud models with the data,
the weights must  
be multiplied by the conversion factors.  
 
\end{table*}  

The modelling of the near- and mid-infrared continuum of a selected
sample of
Seyfert galaxies presented by Alonso-Herrero et al. (2003) is
presented in this section.  The continuum data provided by
these authors has  been subtracted from the light contribution of
the background galaxy and  are thus  the best available data for analysis
of the most central AGN emission.
For modelling purposes, those  
objects presenting the widest wavelength coverage and minimum number
of upper limits were selected.  For NGC 1068, the dataset from Rieke \&
Low (1975) is also included
(open symbols).
Objects which were previously
modelled in detail are omitted :
NGC 4151 (Contini et al 2002), NGC 4388 (Ciroi et al 2003),
NGC 5252 (Contini et al. 1998),
and  NGC 4051 (Contini \& Viegas 1999).

In this paper, we try to explain all the galaxies with
prototype models.  In all cases,   clouds with \Vs = 100~ \kms are
present and in most of them \Vs = 500~ \kms clouds with different d/g
also contribute to the fit.  In a few cases, higher velocity
clouds are  also required.

The results produced by  multi-clouds models are shown in Figs. 5 and
6 for Seyfert 1.8-2 and Seyfert 1.5-1, respectively.  The input
parameters of the single-cloud models are listed in Table 1. For each
galaxy, the single-cloud models used to fit the continuum are given in
Table 2. The covering factors  of the clouds are related
to the relative weights  (see Table 2)
which are used to obtain the   best fitting multi-cloud model.
Models M1 and M7 (Table 1) contribute significantly  
to  the fitting of the SED   in
the near-IR and soft X-rays ranges, respectively. Their
continua are shown in Fig. 7. Model M3 is already  presented in Fig. 1.

To illustrate the results of multi-cloud models, in the panels
corresponding to UM 148, NGC 3786, Mkn 335, and Mkn 530 , the
single-model contributions are summed up for the dust emission and for
the free-free emission separately and the best fitting SED is shown as
a thick solid line in Figs. 5 and 6.  For the other galaxies the
domains of the different models are well separated, so only
single-cloud models appear in the figure. The summed SED would not
change the picture.

We note that in most cases,  the data at the lowest frequency, 16 \mum,  
should be taken  as  an
upper limit, as it is largely contaminated by the host galaxy contribution  
due to the large ISO aperture. Therefore, the d/g for these clouds,
which usually are associated  with velocities of about 500 km/s,  is
 poorly constrained.
In two galaxies, Mkn 334 and NGC 3362, the flux at
optical frequency is also an upper limit.
In NGC 7479 large d/g values ($>$ 8 10$^{-12}$) are
indicated by the data, in agreement with the upper limit in the
optical domain.  Data at higher IR  frequencies are needed to
confirm this d/g  value.

In brief, the infrared continuum of all the galaxies can be modelled
by multi-cloud models obtained by the weighted sum of different
single-cloud models.  A fine tune of these results can be obtained by
further modelling the emission-line spectra and the overal SED.
Nevertheless, from the detailed analysis
of a representative number of Seyfert galaxies (Contini et al.  1998a,b,
2002, 2003) no dramatic changes for the results of the  IR-optical  SED are
expected.  

A comparison between Seyfert 1.8 -- 2 and Seyfert 1.5 -- 1 modelling  shows
that IR-optical continuum fluxes are generally higher in Seyfert
1.5 -- 1.0; moreover, the flux tends to decrease toward higher
frequencies for some Seyfert 2 galaxies.  That behaviour is well
followed by the models.  For all
types of galaxies, the continuum between 3 $\mu$m and 10 $\mu$m is
mainly due to thermal bremsstrahlung by gas with relatively
high density and heated by a low velocity shock  (model M1). This
indicates that dust emission in this region is relatively low, in
contrast with most of the luminous IR galaxies where reemission by
dust dominates at wavelengths even lower than 10 $\mu$m (Contini \&
Contini 2003).   Clouds with \Vs $\geq$ 500 \kms contribute to the SED at
higher frequencies.  {\it The relative weights of clouds with \Vs $\geq$
500 \kms
and clouds with \Vs=100 \kms determine the slope of the multi-cloud
continuum SED.}  In average, the ratios of  relative weights
for models M1 and M3 (see Table 2),
which appear in almost all the galaxies, are
higher for Seyfert 1.8 -- 2
(3.7 10$^4$) than for Seyfert 1.5 -- 1 (1.3 10$^4$).
These models correspond to velocities of
100 \kms and 500 \kms, respectively.   M1 is shock-dominated
while  M3 is radiation-dominated.
From the point of view of the unified model for the AGN, it is
expected that
high velocities clouds, reached by a more intense radiation,
should play a more significant role
in the spectra of broad-line objects,  e.g. the NLS1 galaxy Ark 564 
(Contini, Rodrigues-Ardila \& Viegas 2003), since the line of sight
reaches deeper into the central region, where the high velocity
clouds are. Accordingly, M3 appears in 9 out of 16 narrow-line Seyfert
but in 7 out of 8 broad-line objects.

Interestingly, clouds having low \Vs ~and high \n0,  basically model
M1, are present in all the galaxies. Whether these
low velocity  clouds are dusty
or not, could only be found out by modelling their far-IR nuclear
emission, which correspond to radiation by grains at rather low
temperatures. We notice that these  types of clouds are difficult to
trace in the nebular spectrum because they produce  weak lines.
Moreover,  they were not revealed
by previous investigations of  AGN  spectra  because
 complete datasets between 1.1 and 10 $\mu$m were not
available yet.  

The nature of these clouds  is  being discussed in detail  in  the case
of the NLS1  Mrk 766  -  included in the
Alonso-Herrero et al. sample- (Contini, Rodriguez-Ardila \& Viegas 2003,
in preparation).
An analysis of the far-IR SED is therefore needed to determine the d/g
and thus better constrain the modelling. An exmaple of modelling the
complete SED is given in the Appendix, where
an analytical method
to obtain a first  representation  of  the far-IR to UV
SED of Seyfert galaxies based on our model approach is presented.
For this, a more complete grid of models is given.  

\begin{figure*}
\includegraphics[width=42mm]{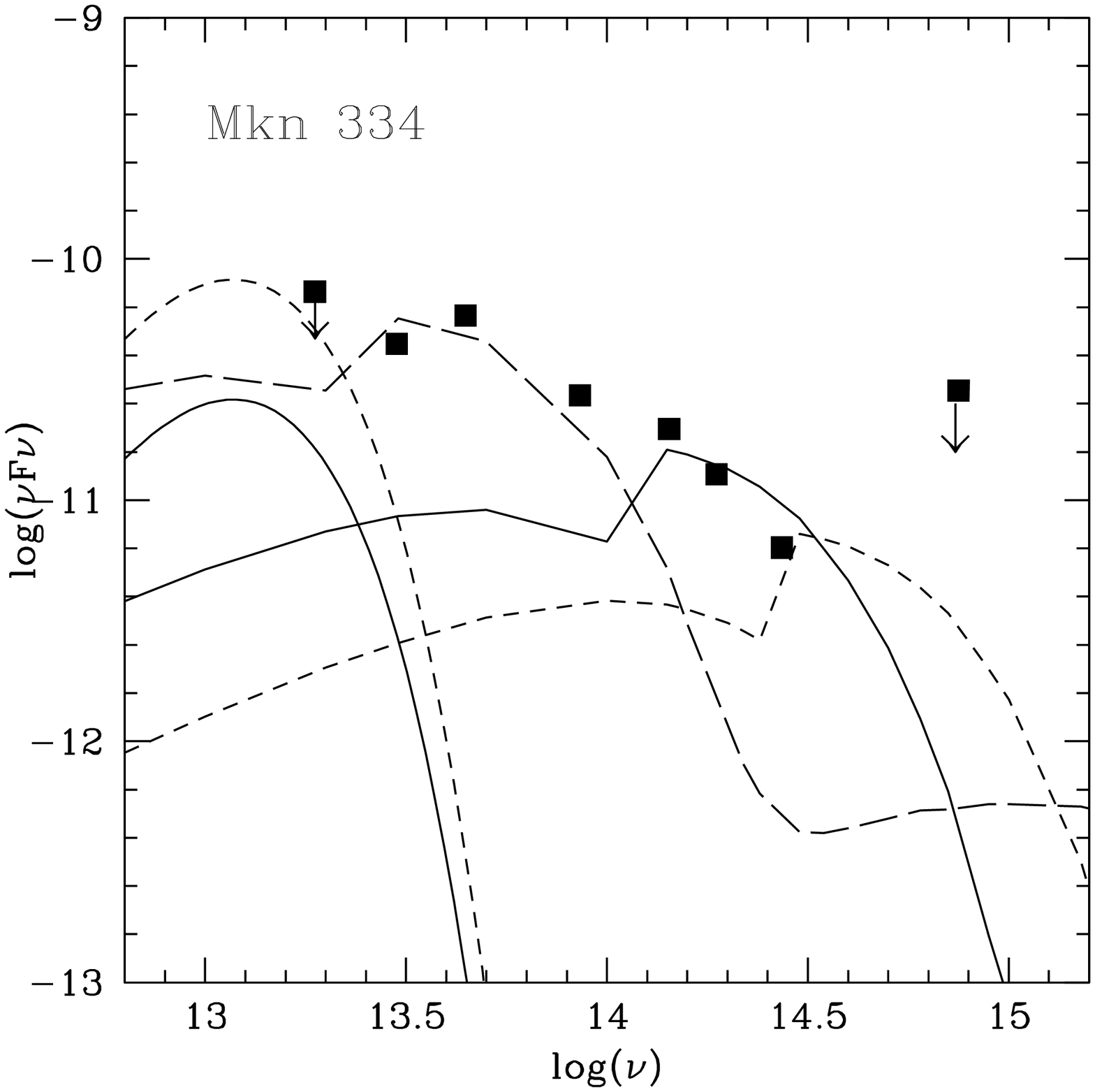}
\includegraphics[width=42mm]{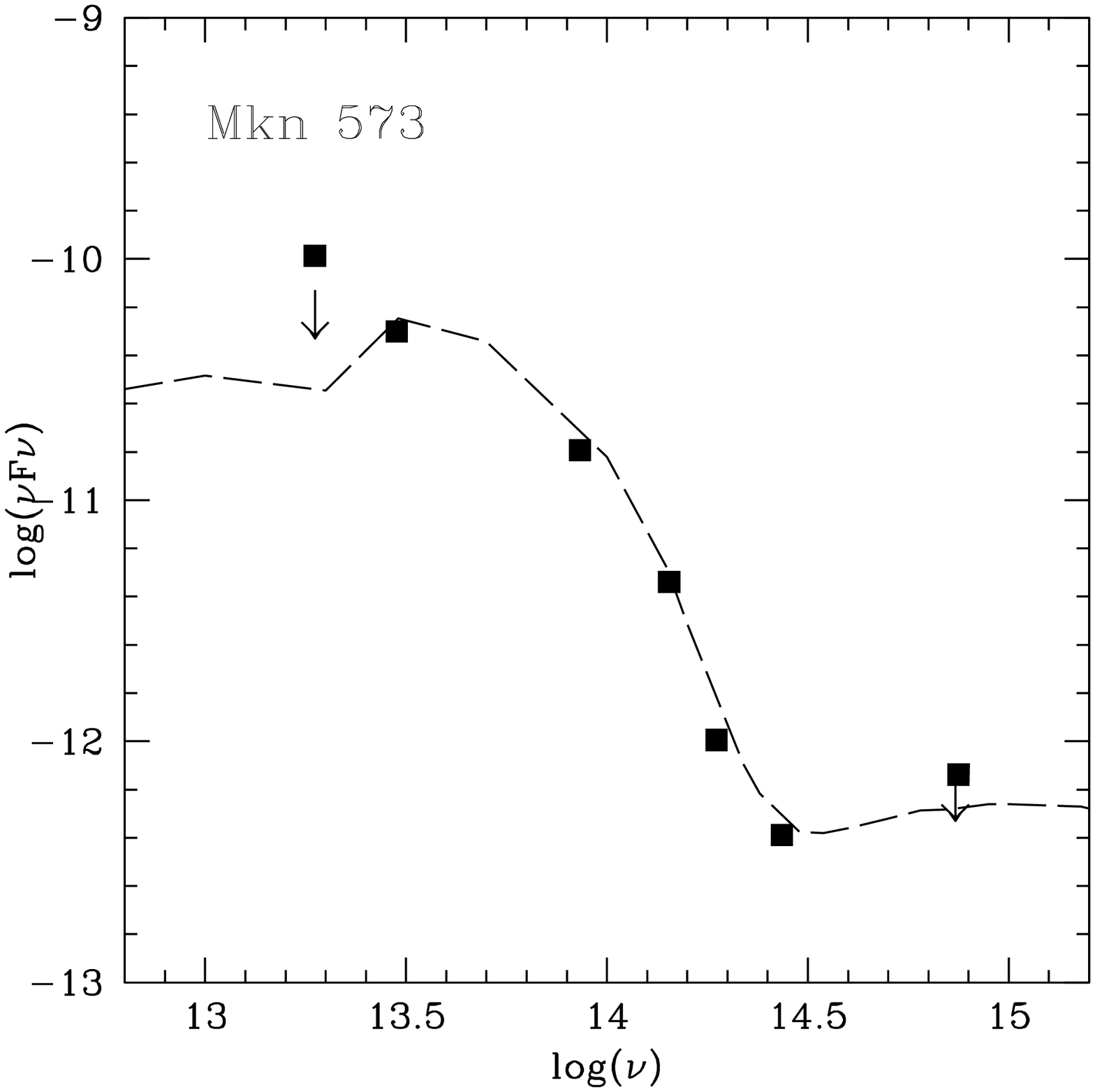}
\includegraphics[width=42mm]{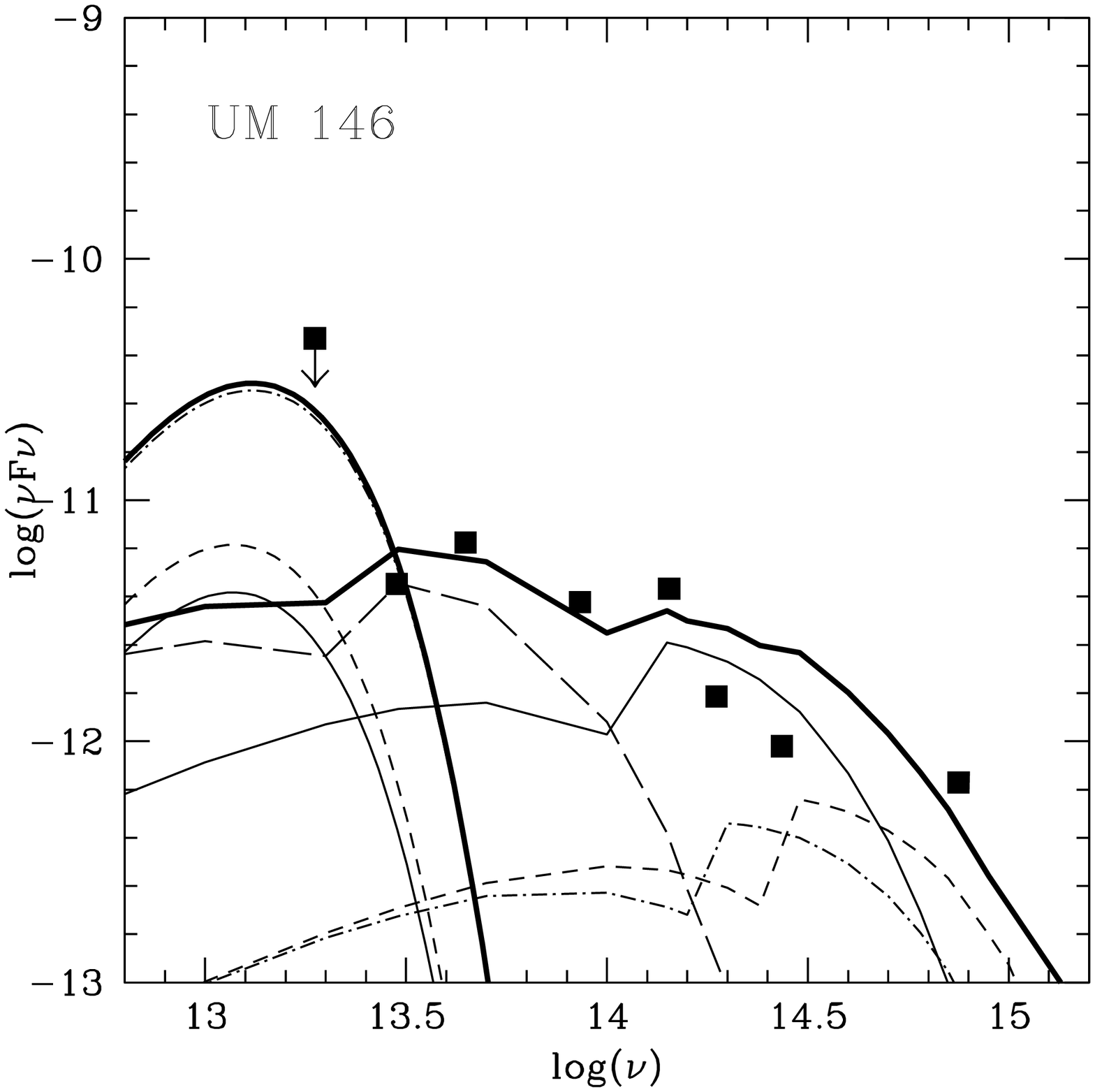}
\includegraphics[width=42mm]{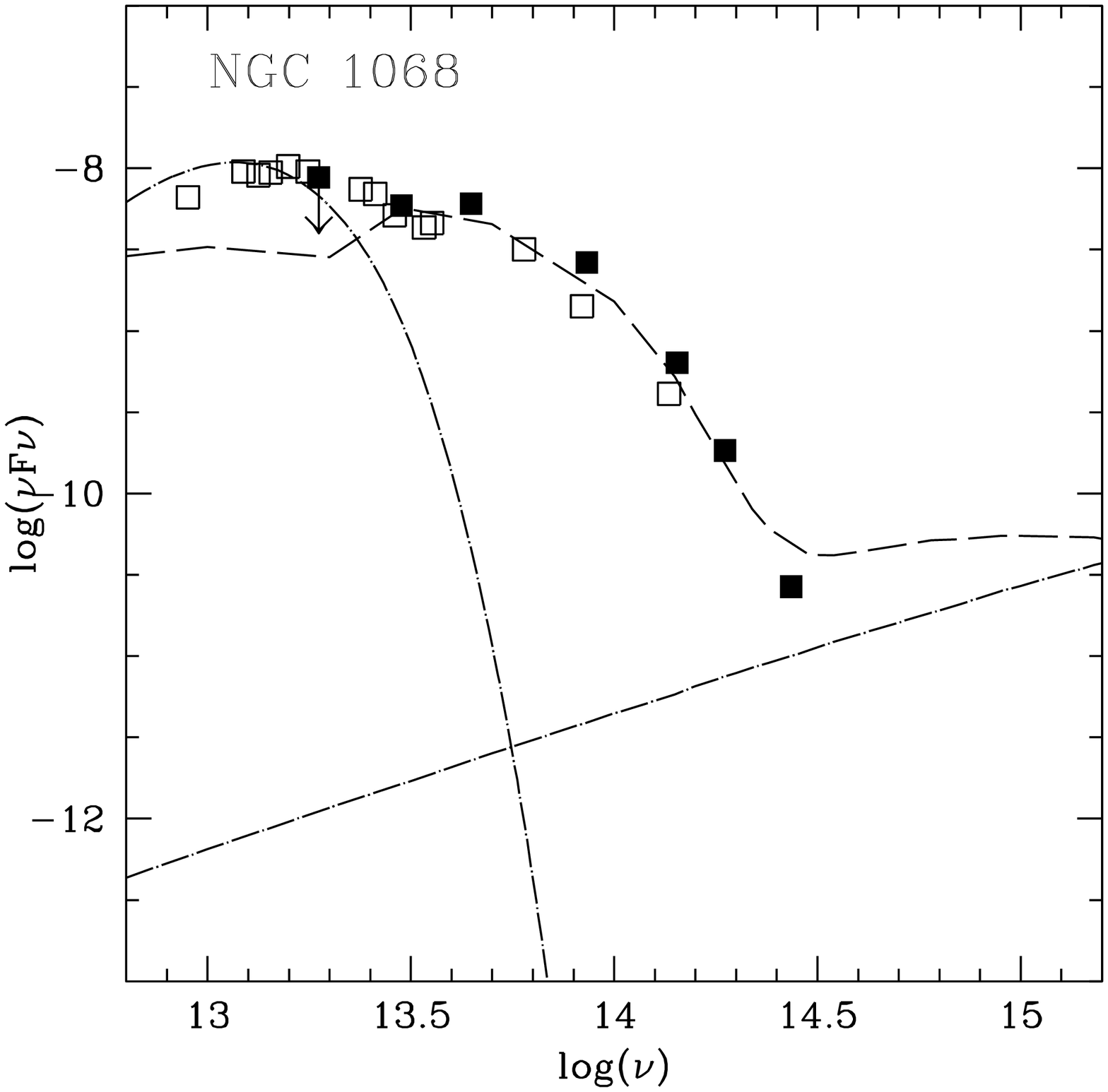}
\includegraphics[width=42mm]{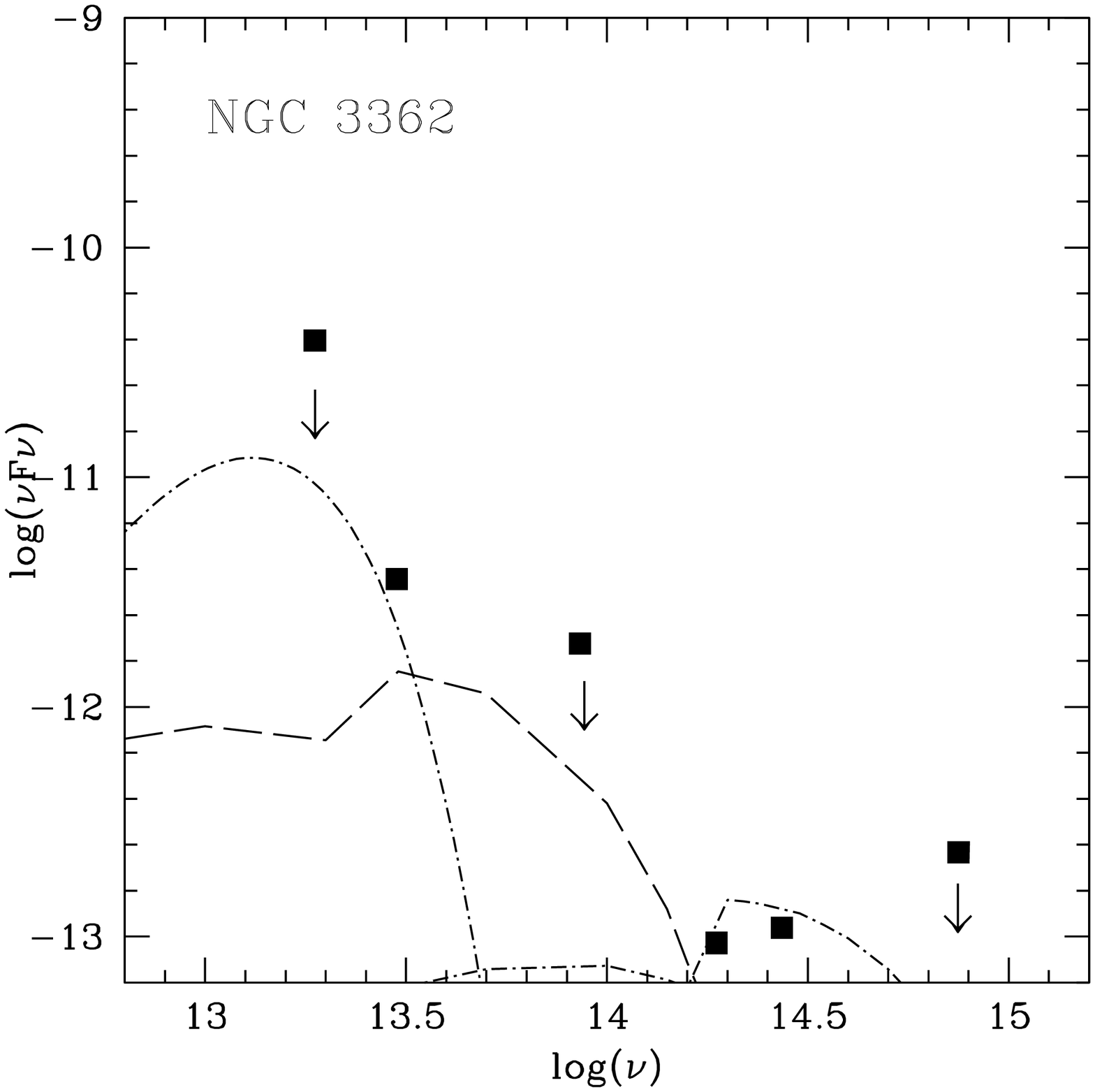}
\includegraphics[width=42mm]{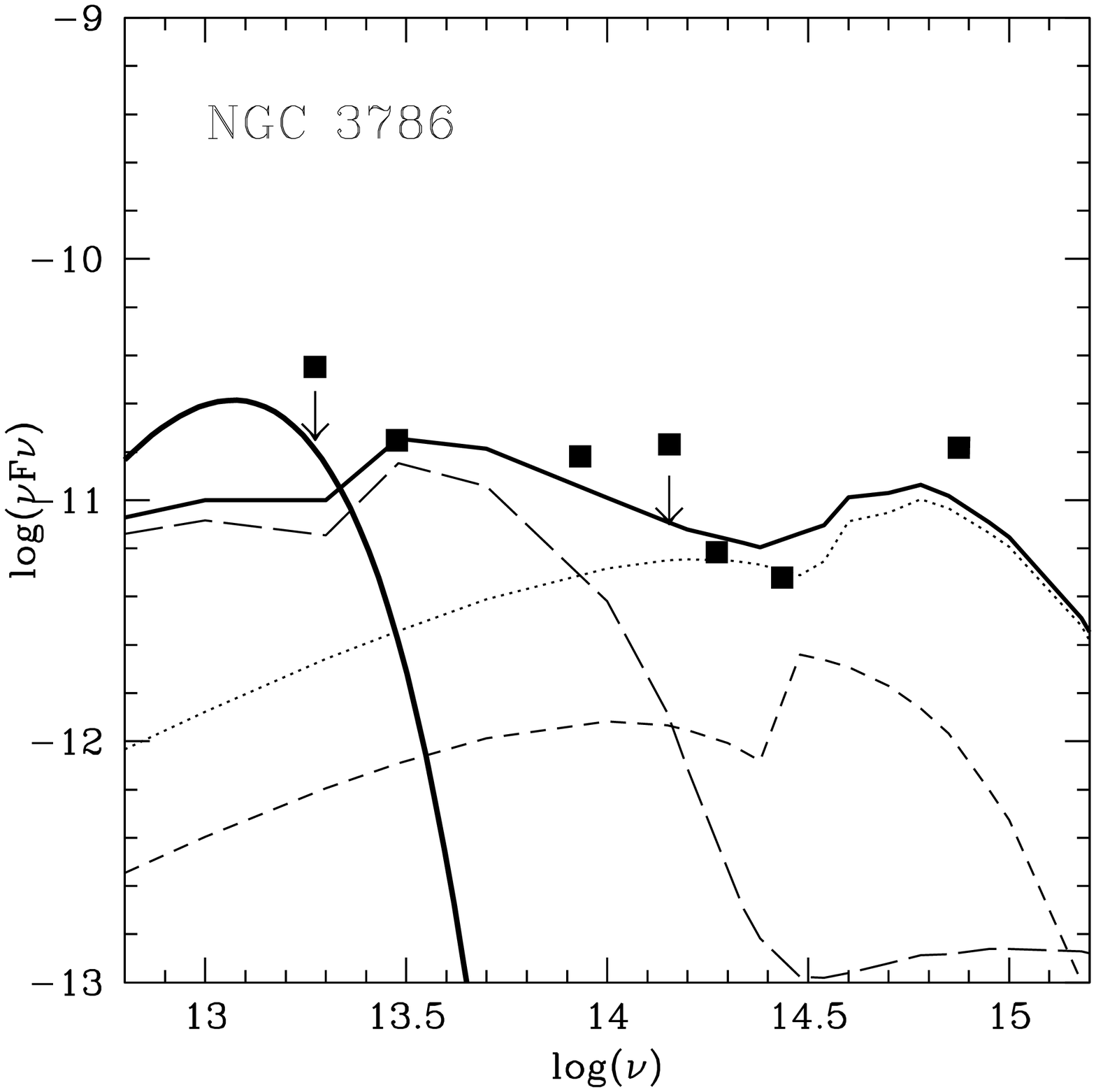}
\includegraphics[width=42mm]{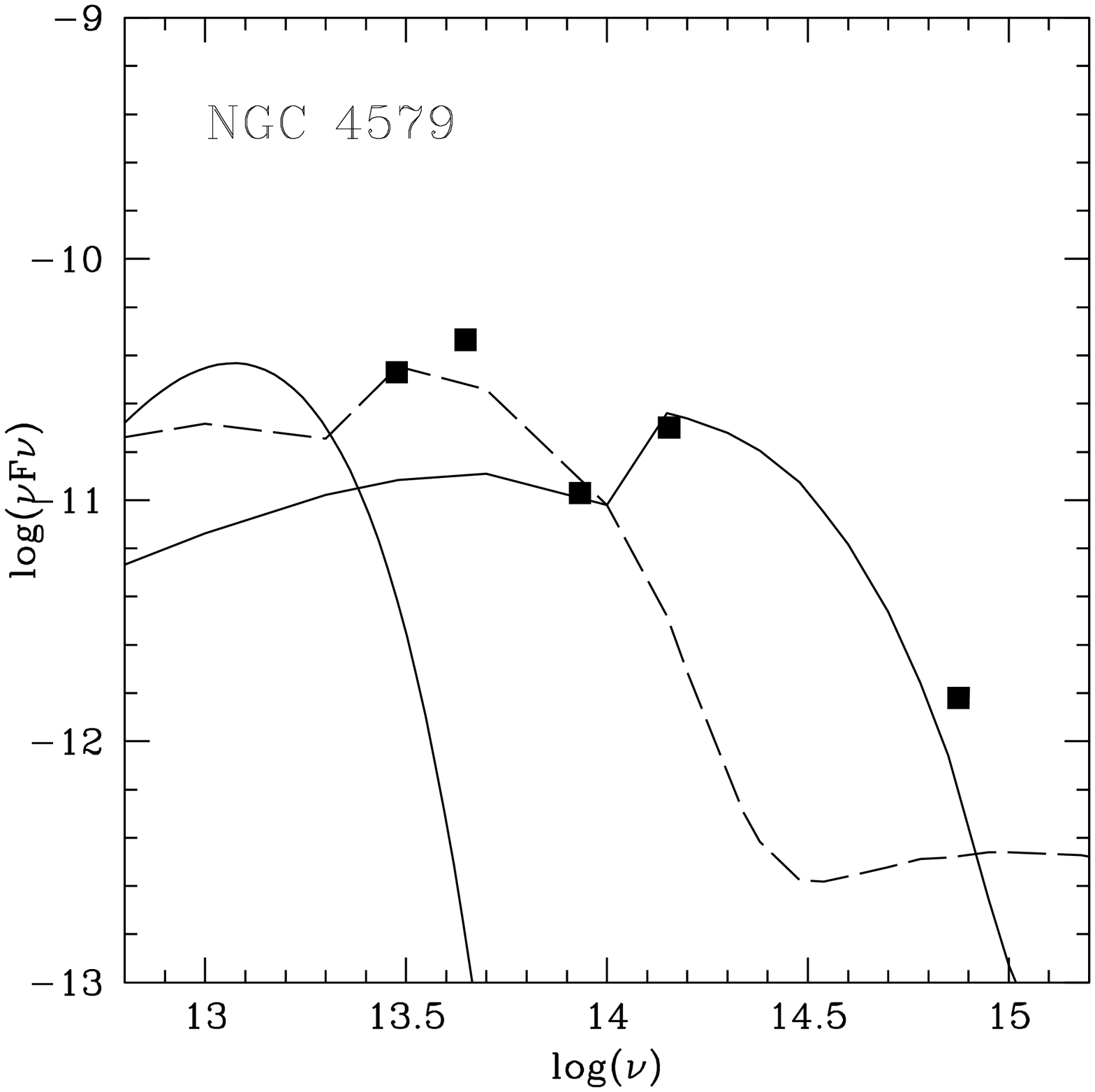}
\includegraphics[width=42mm]{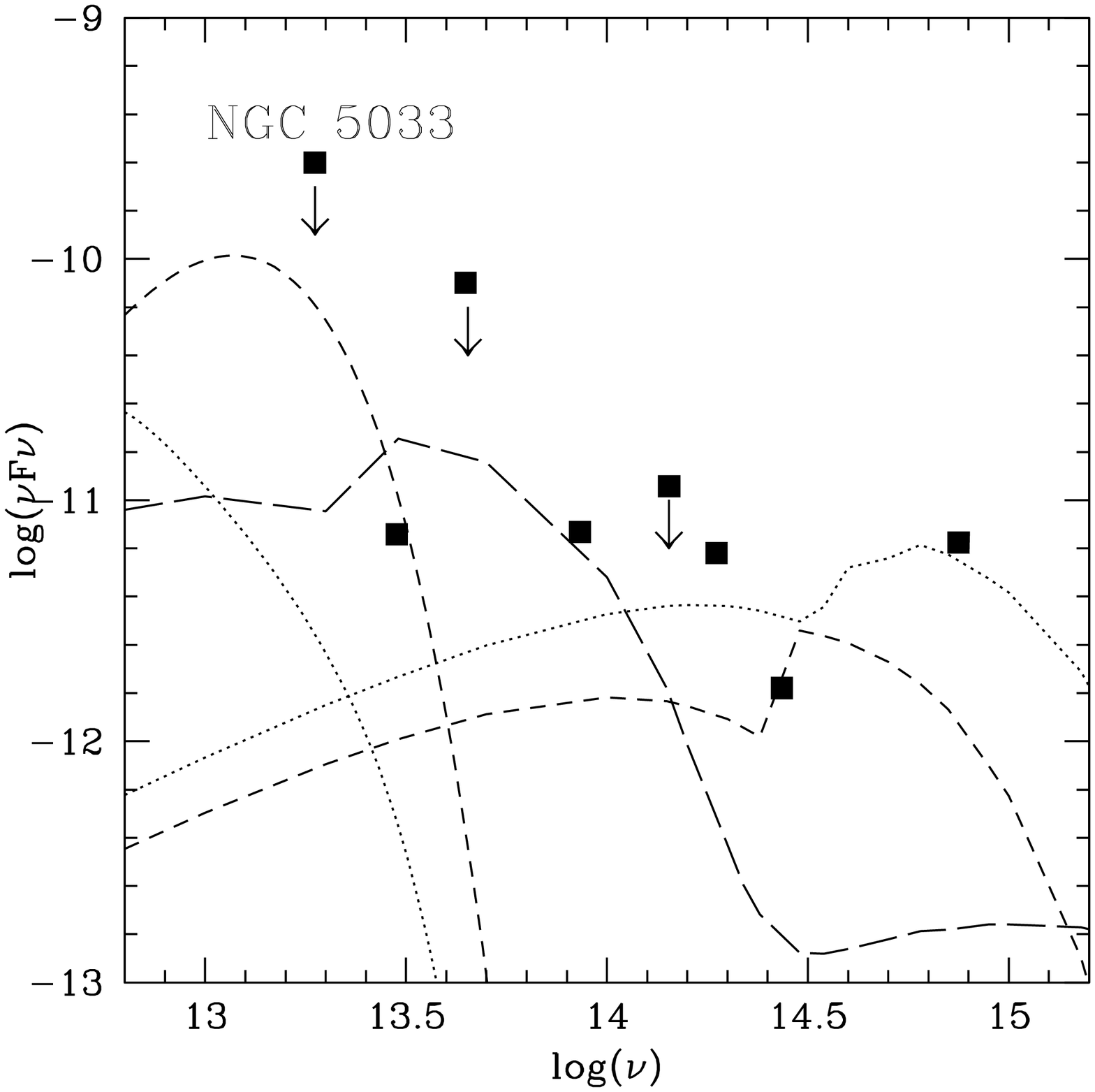}
\includegraphics[width=42mm]{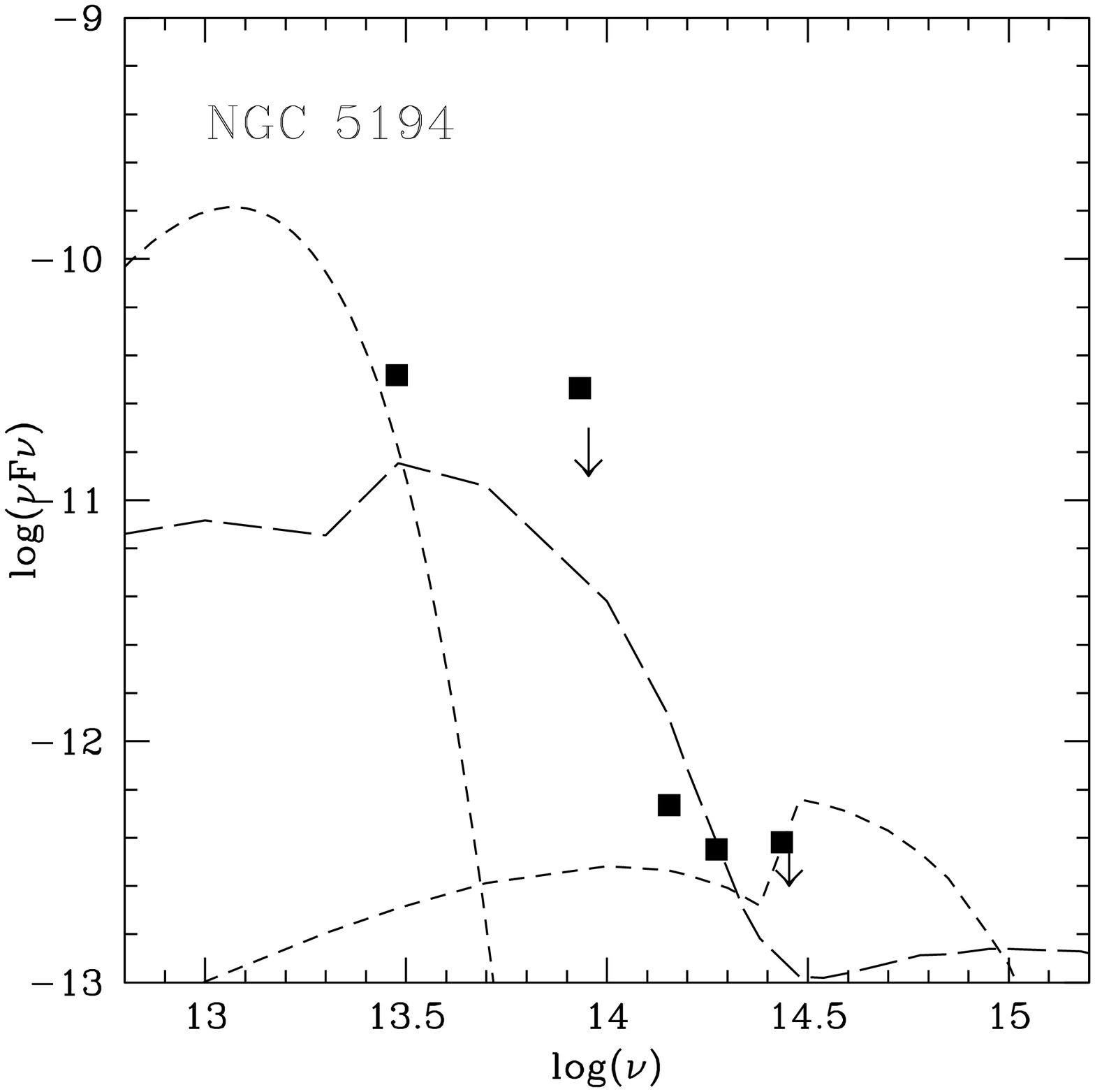}
\includegraphics[width=42mm]{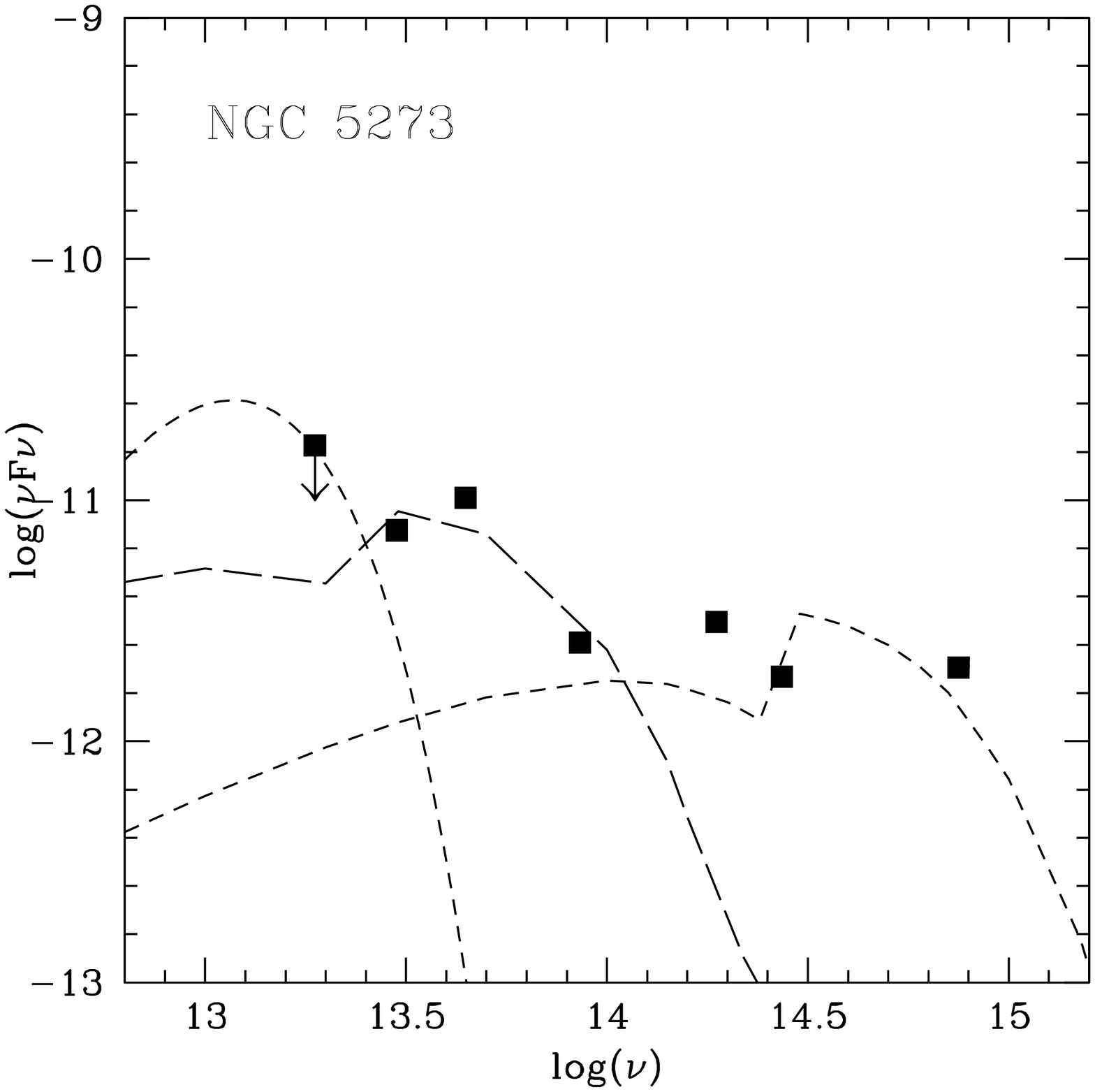}
\includegraphics[width=42mm]{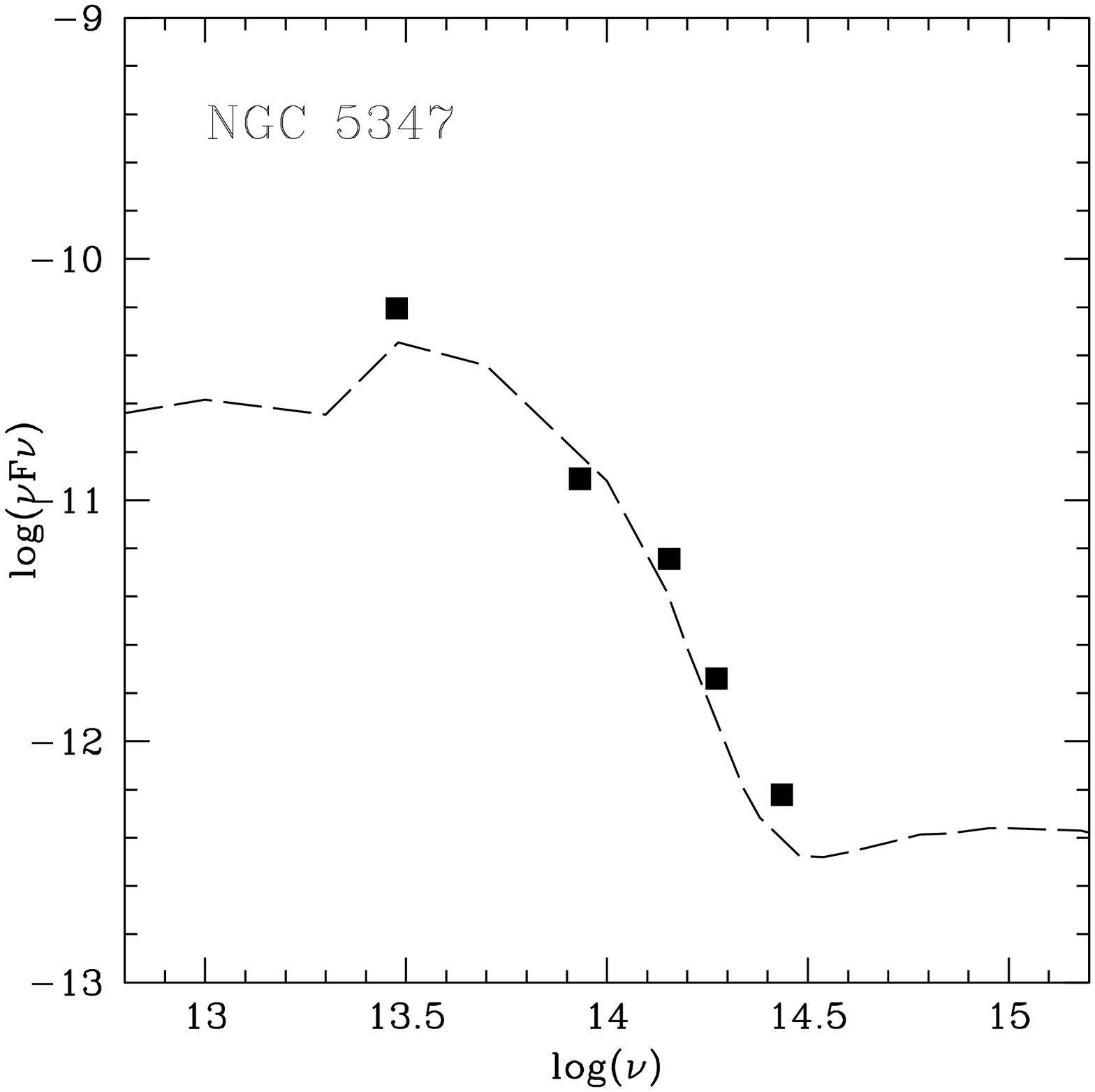}
\includegraphics[width=42mm]{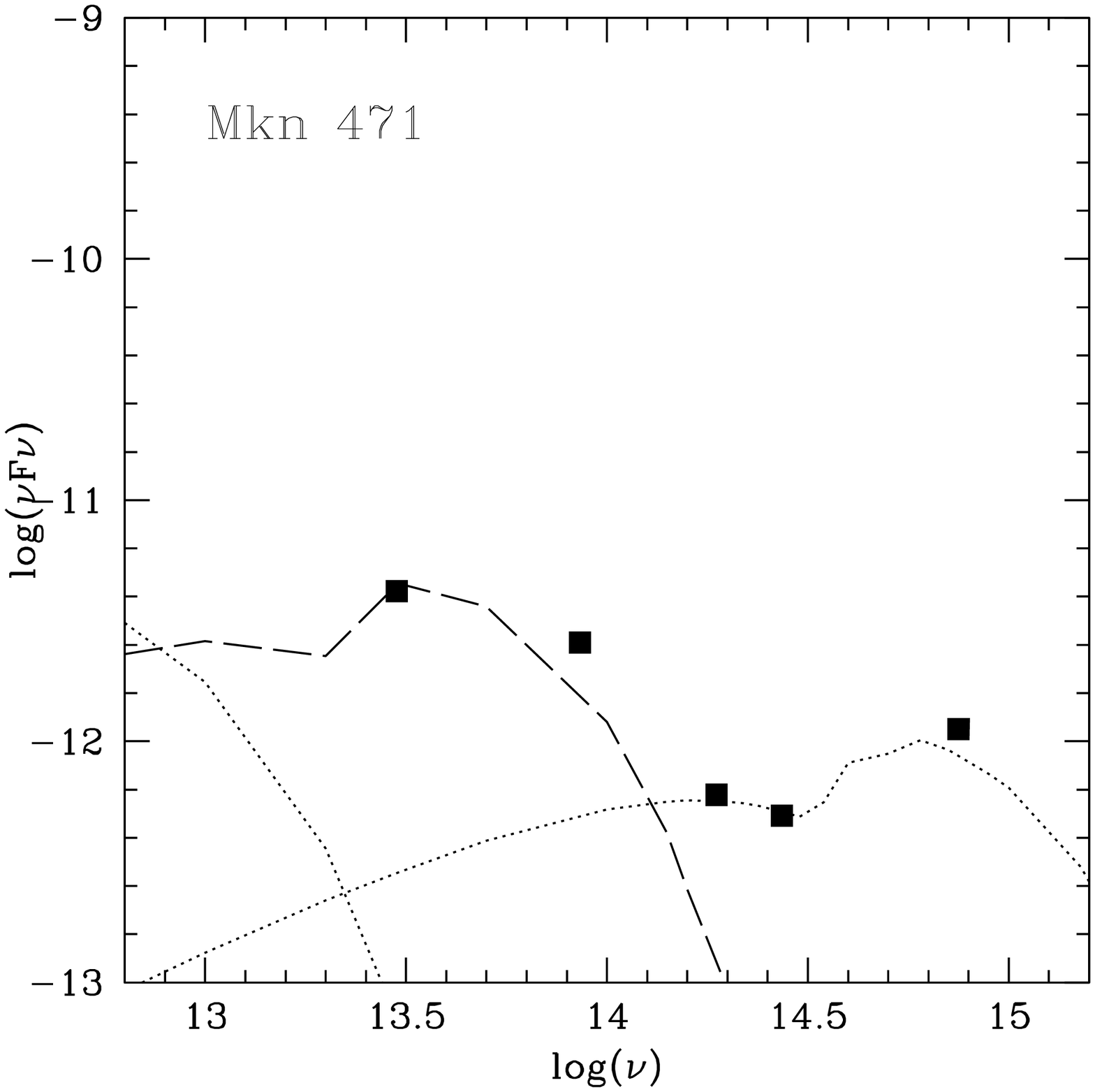}
\includegraphics[width=42mm]{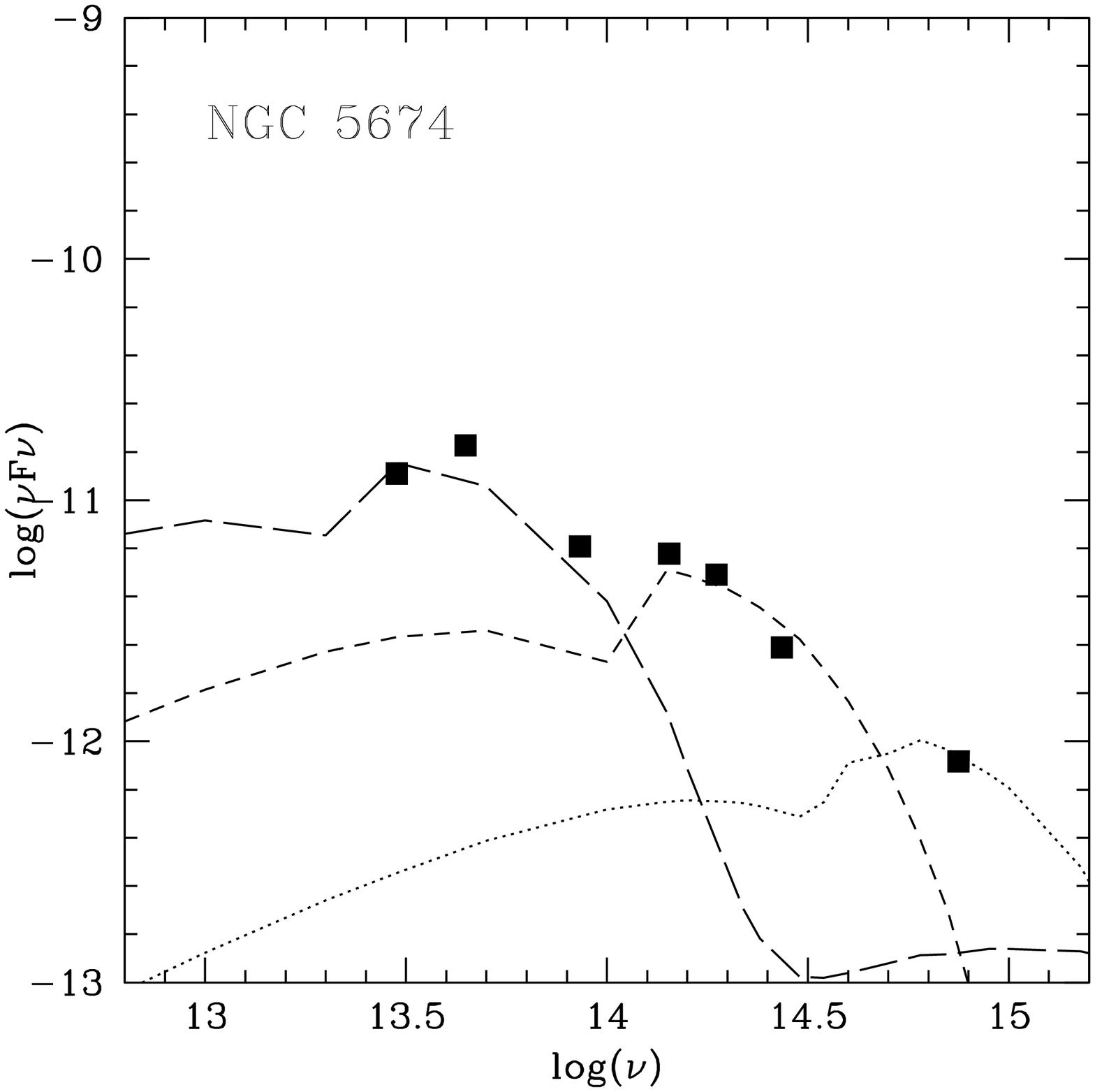}
\includegraphics[width=42mm]{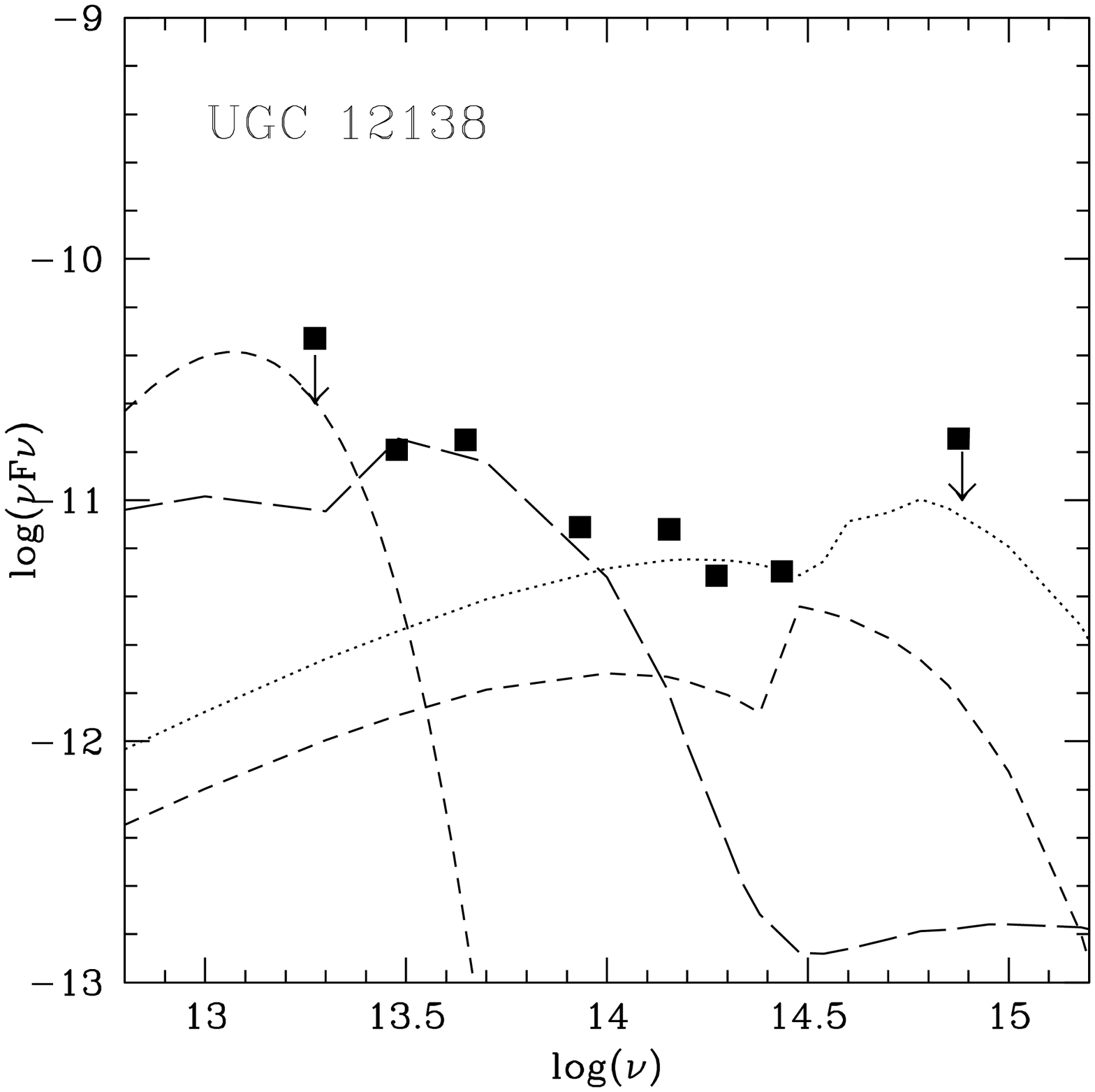}
\includegraphics[width=42mm]{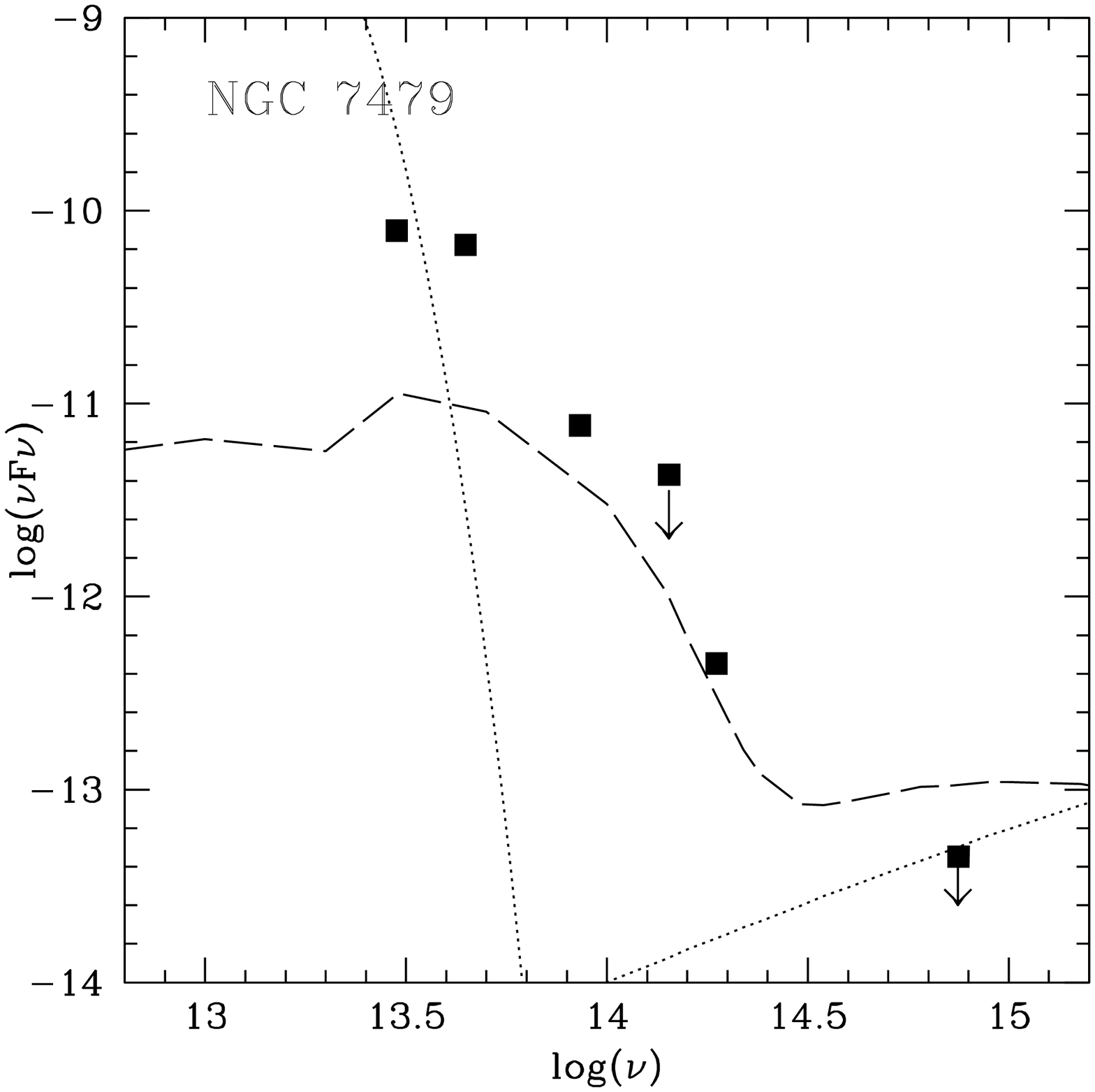}
\includegraphics[width=42mm]{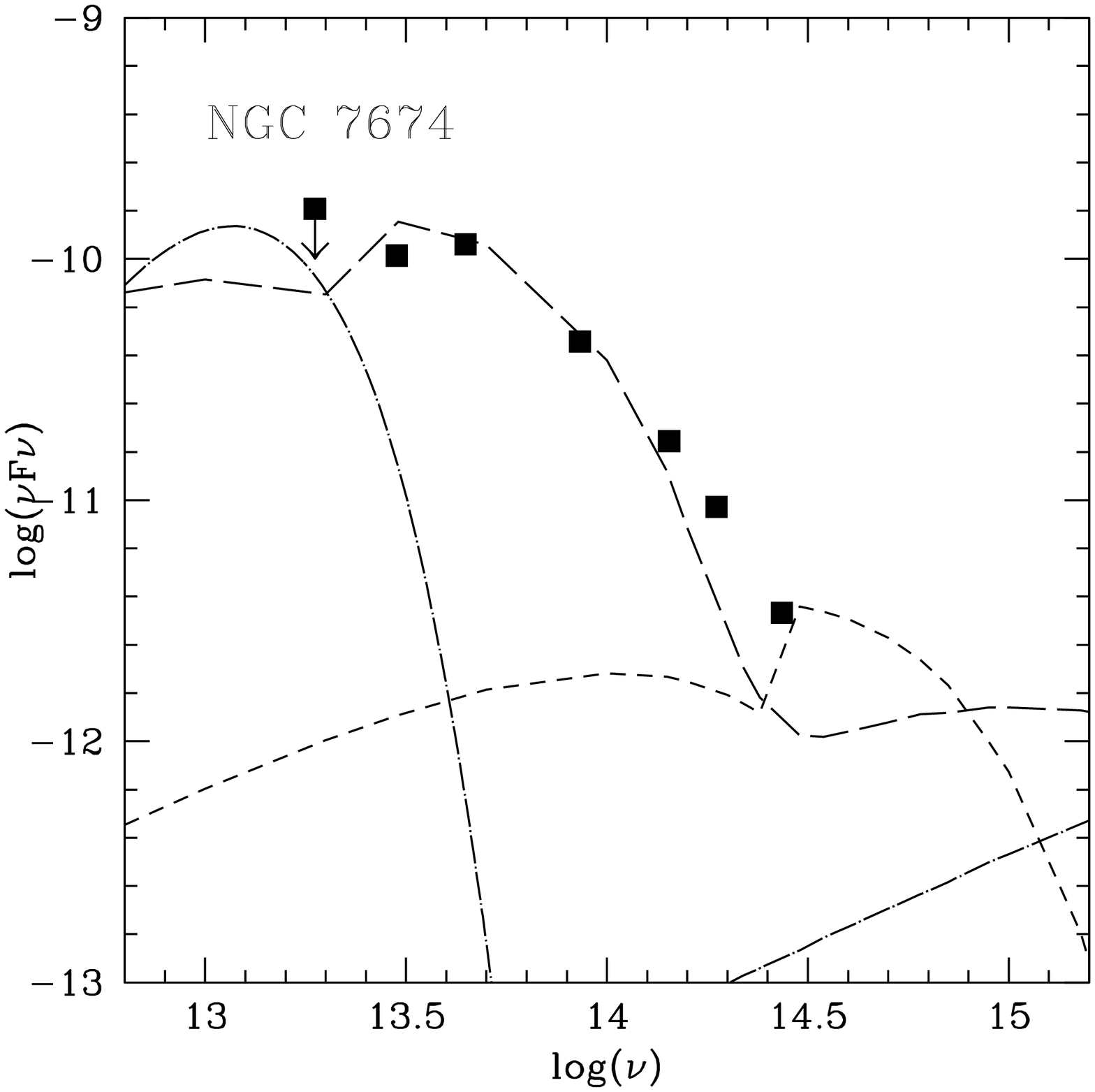}
\caption
{The modelling of selected galaxies from Alonso-Herrero et al. sample:
Seyfert 1.8-2
}
\end{figure*}

\begin{figure*}
\includegraphics[width=42mm]{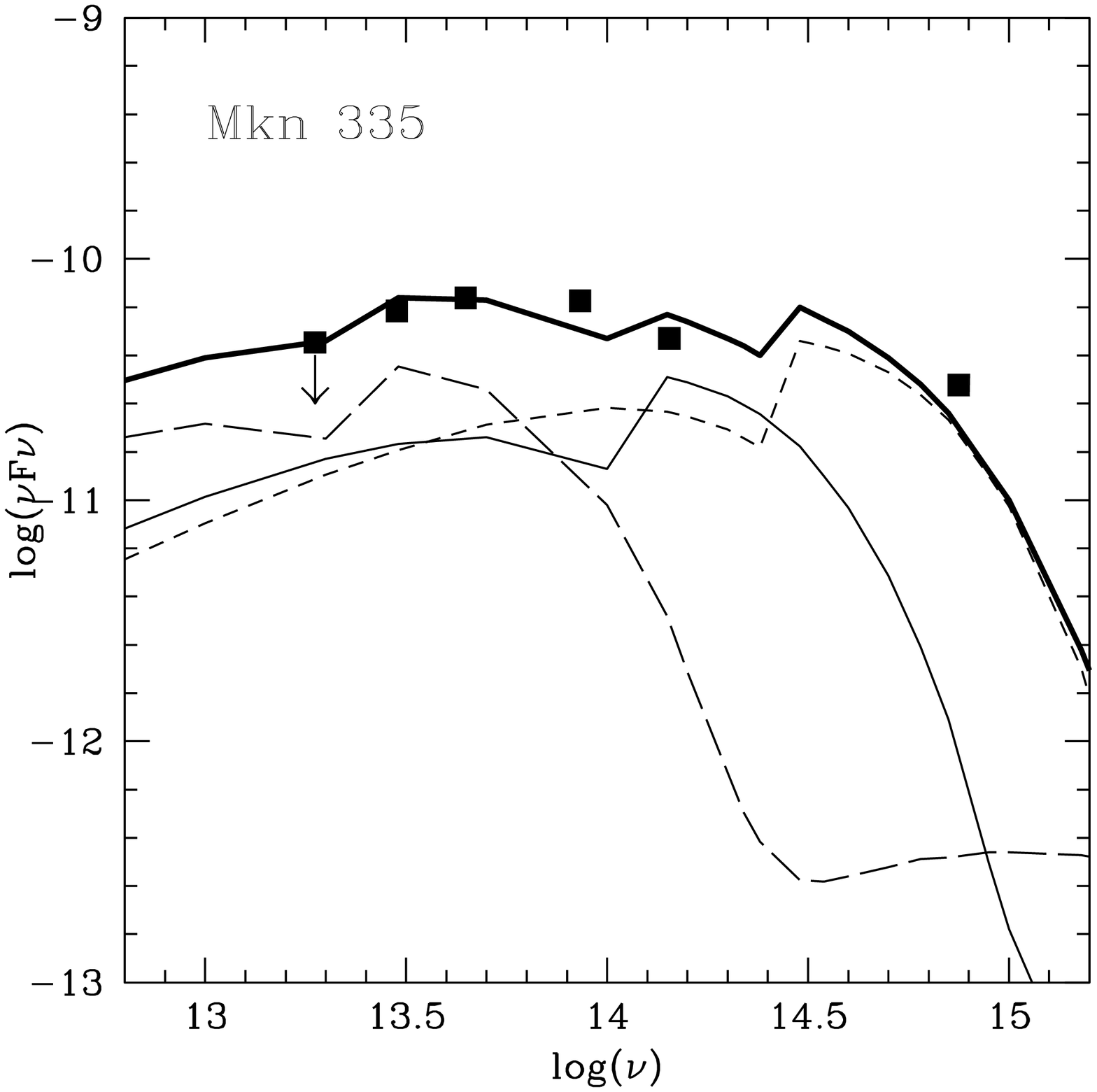}
\includegraphics[width=42mm]{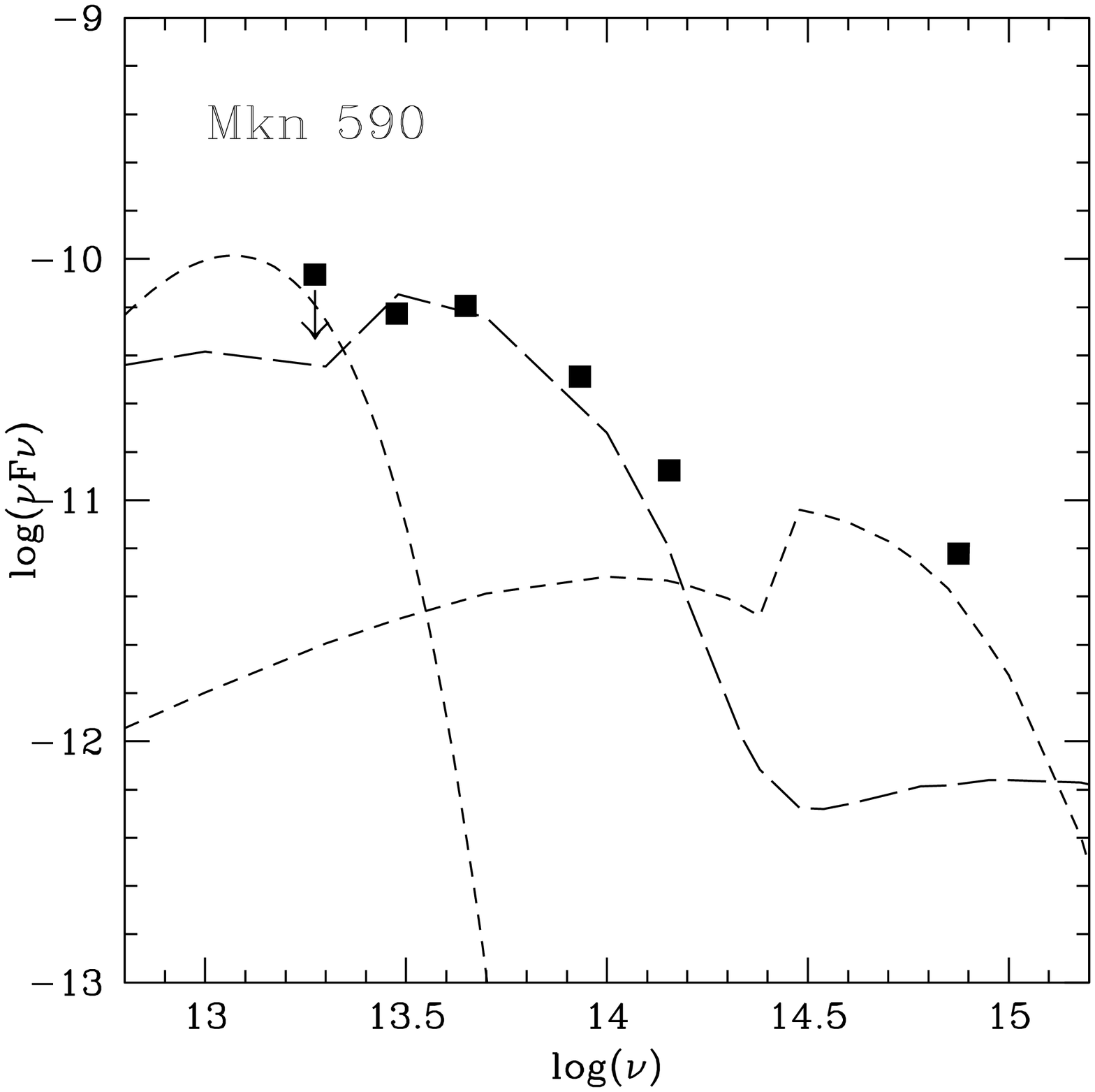}
\includegraphics[width=42mm]{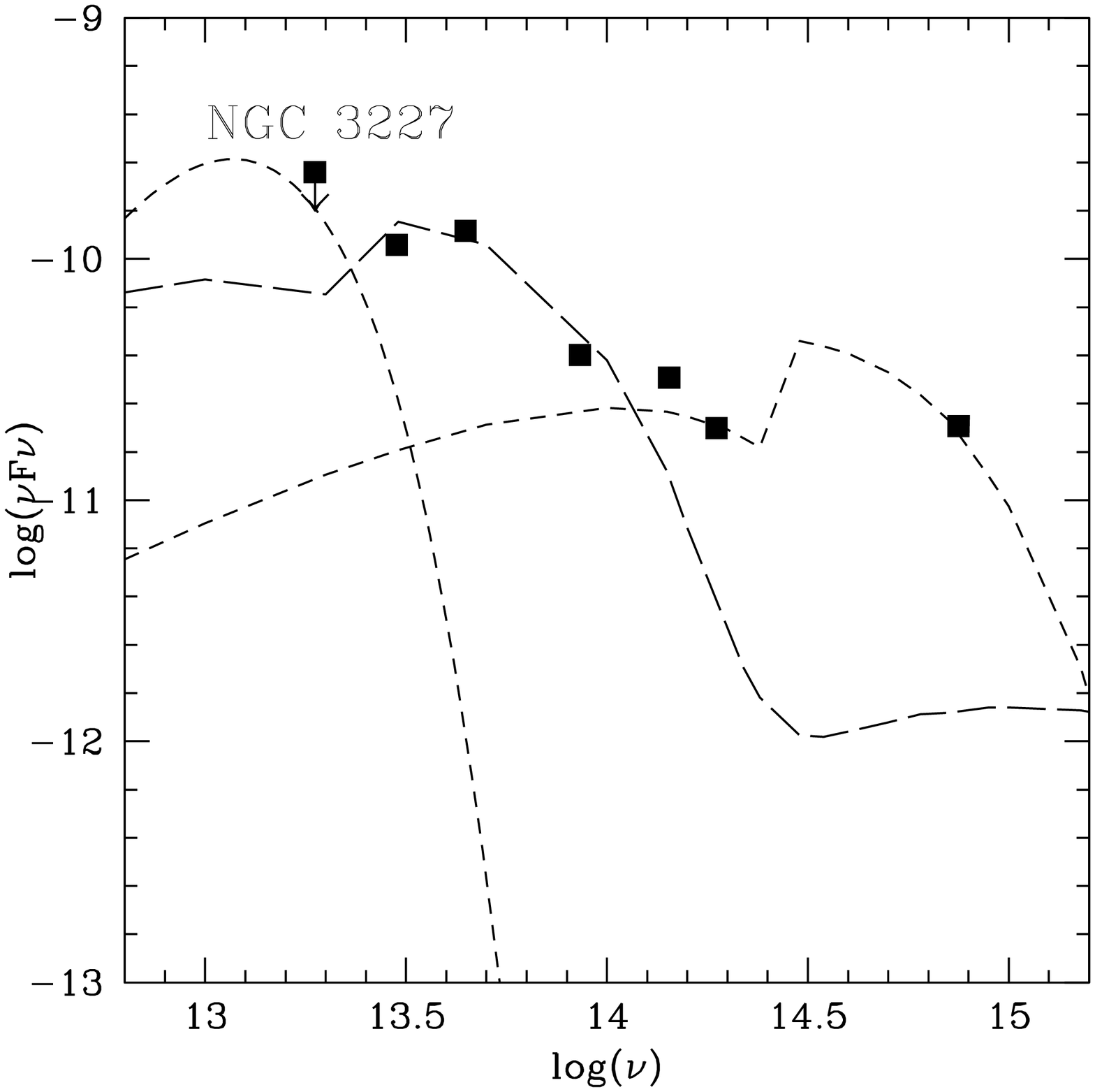}
\includegraphics[width=42mm]{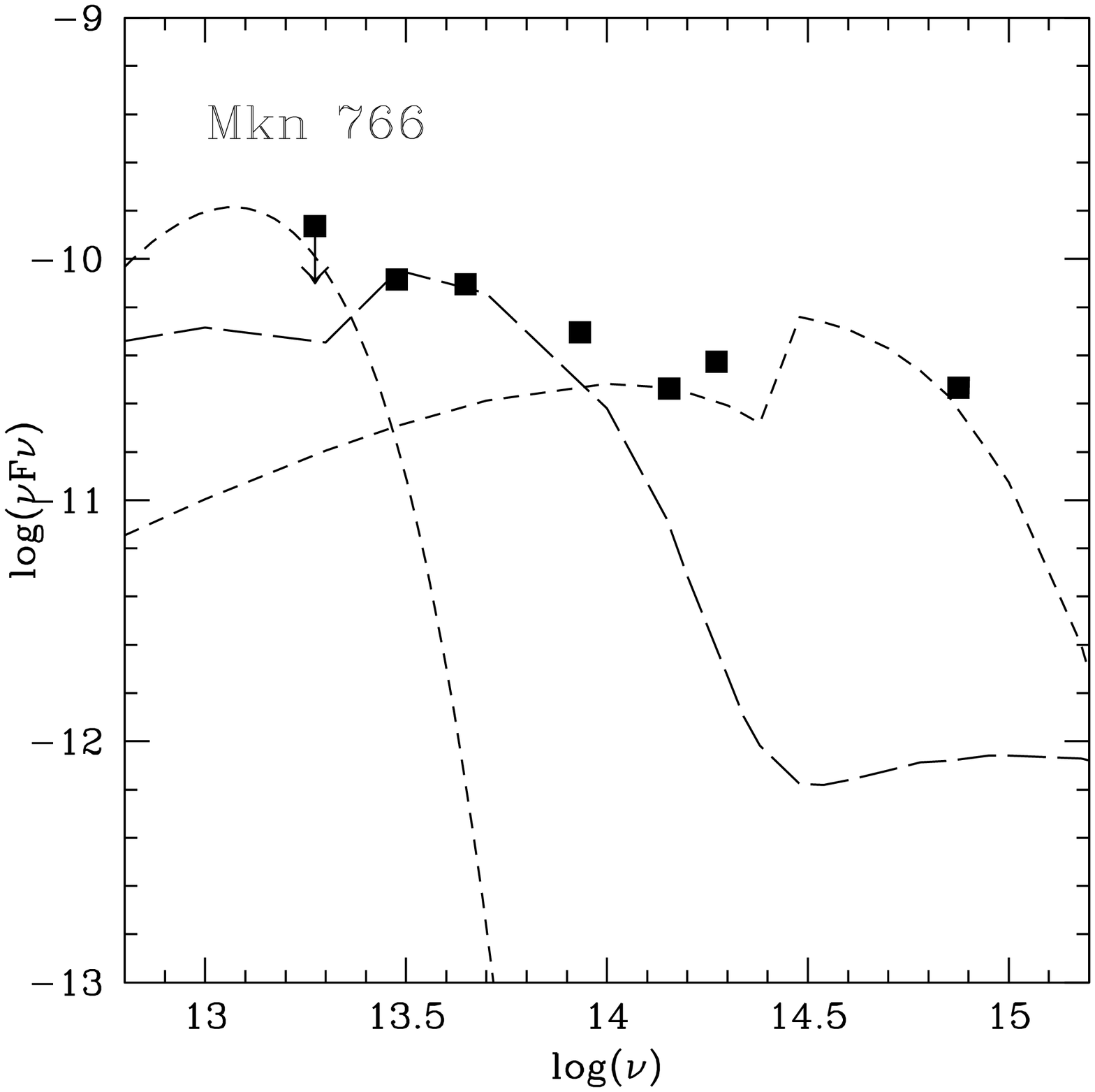}
\includegraphics[width=42mm]{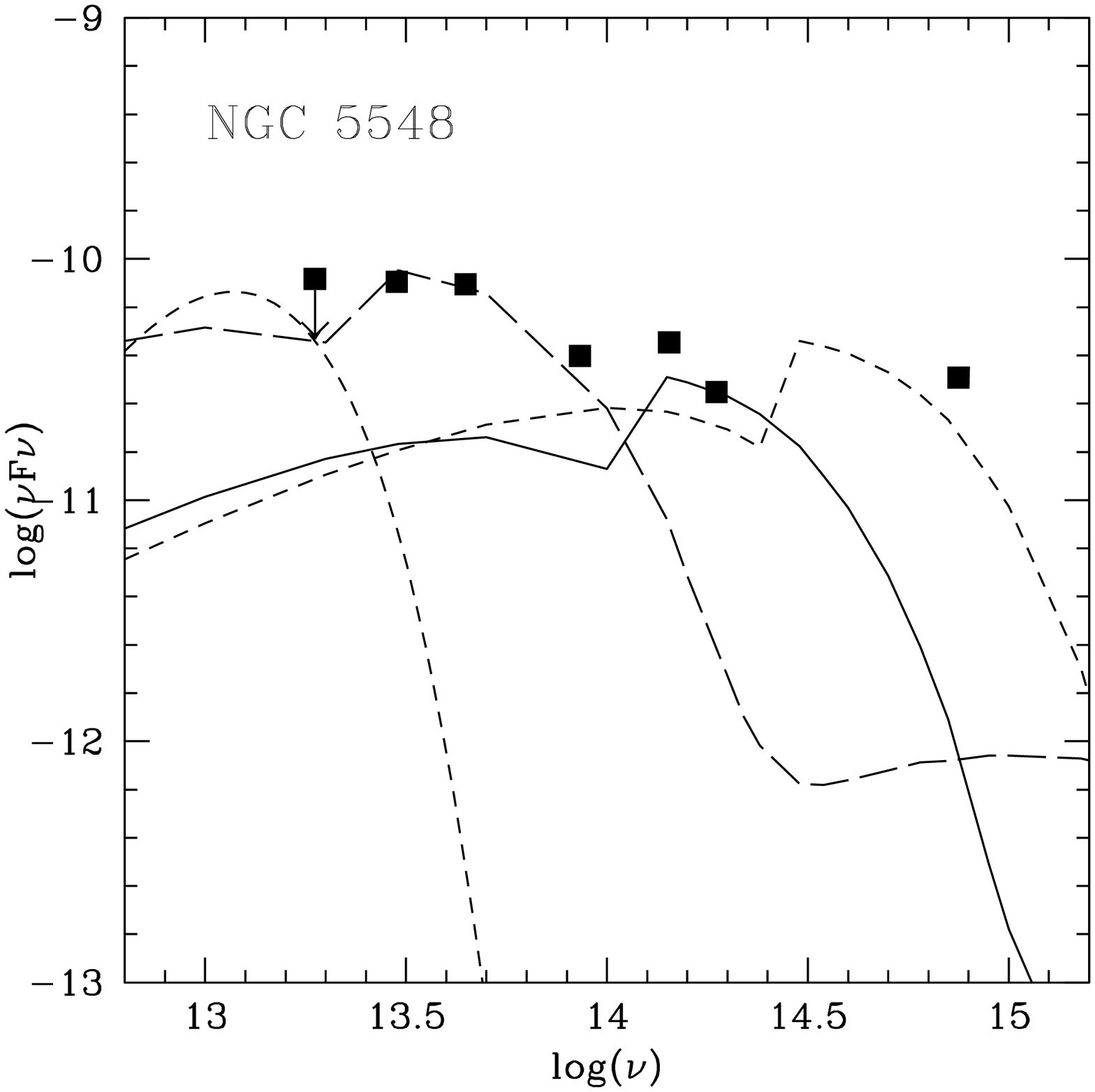}
\includegraphics[width=42mm]{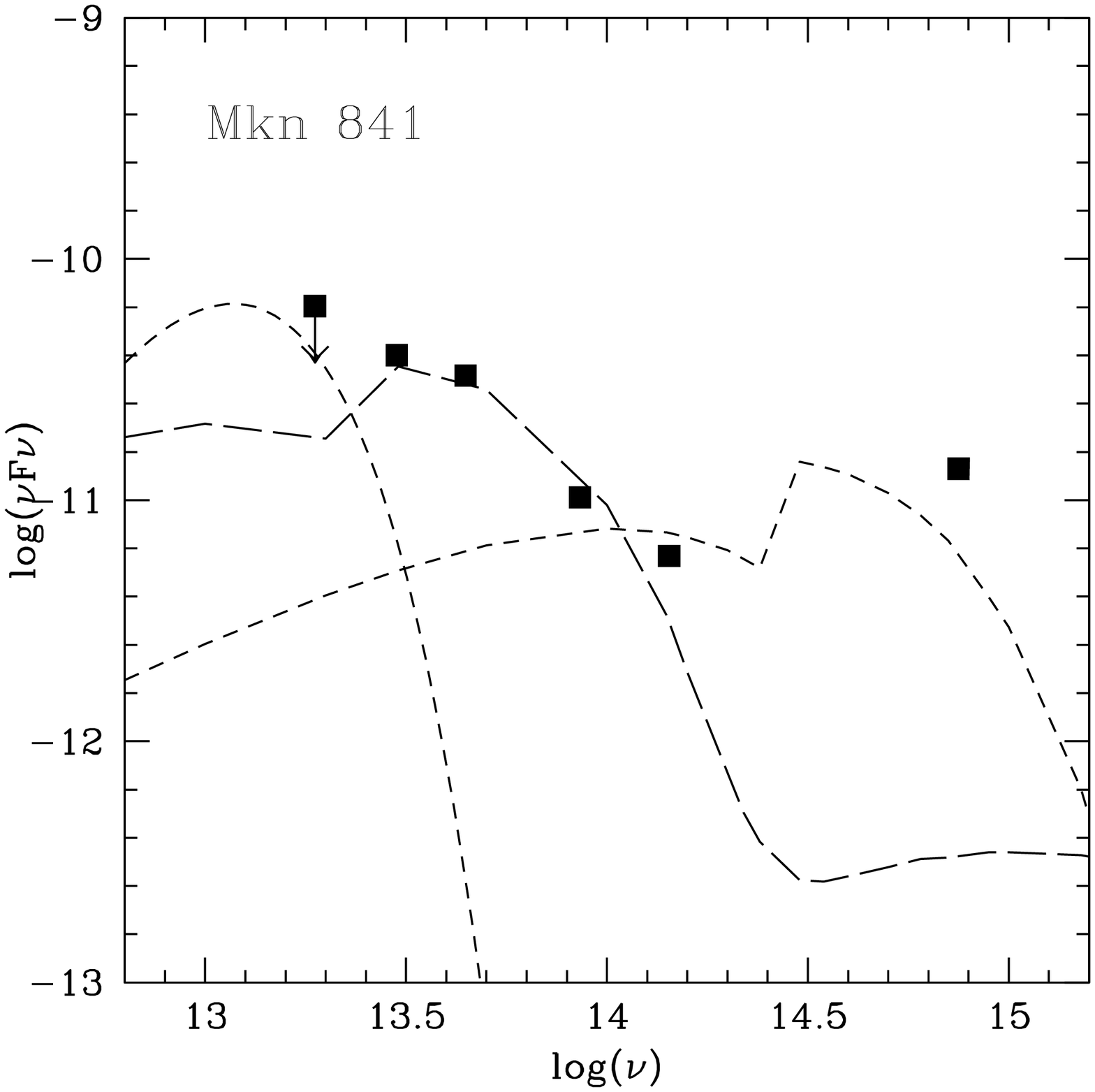}
\includegraphics[width=42mm]{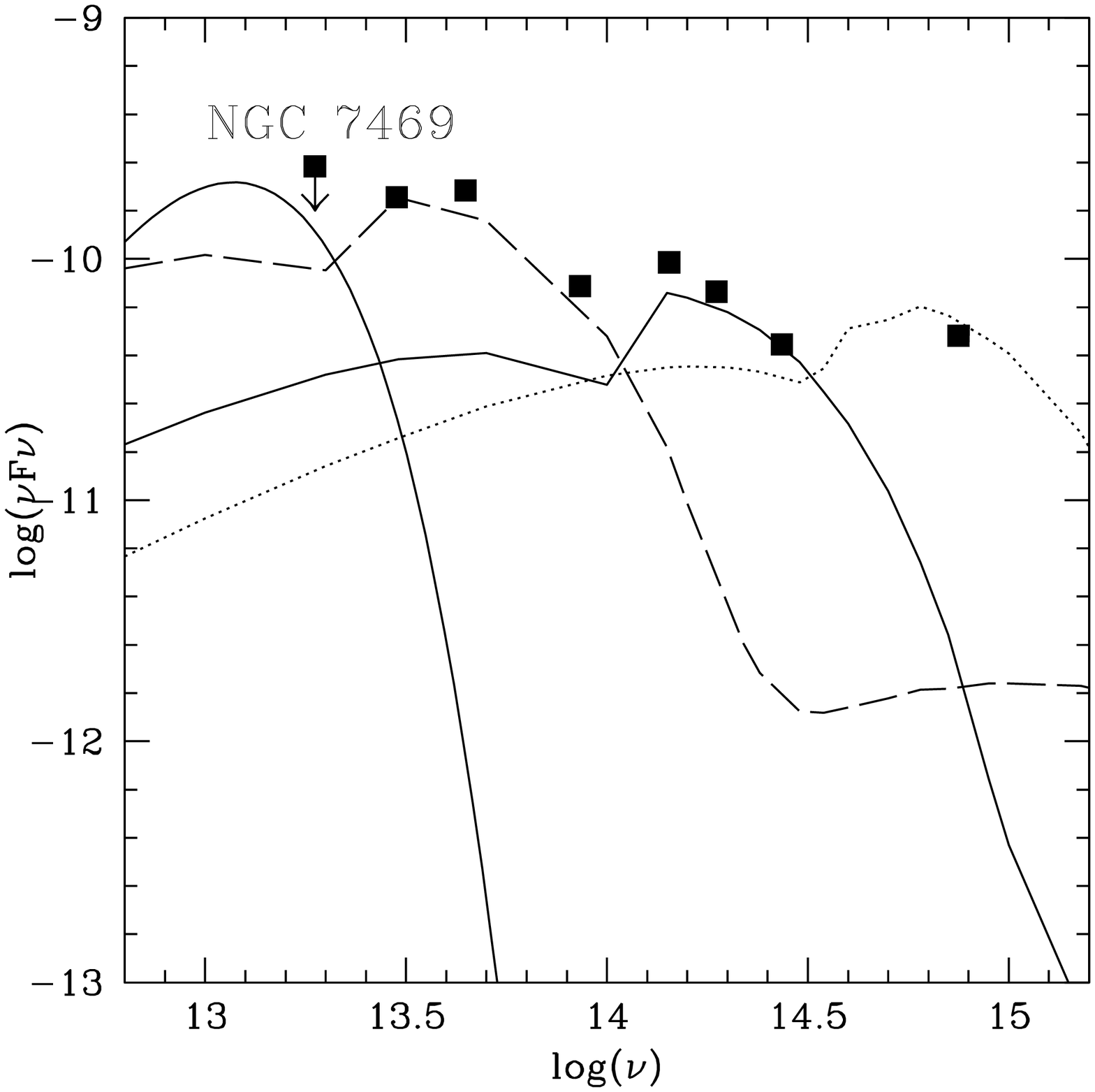}
\includegraphics[width=42mm]{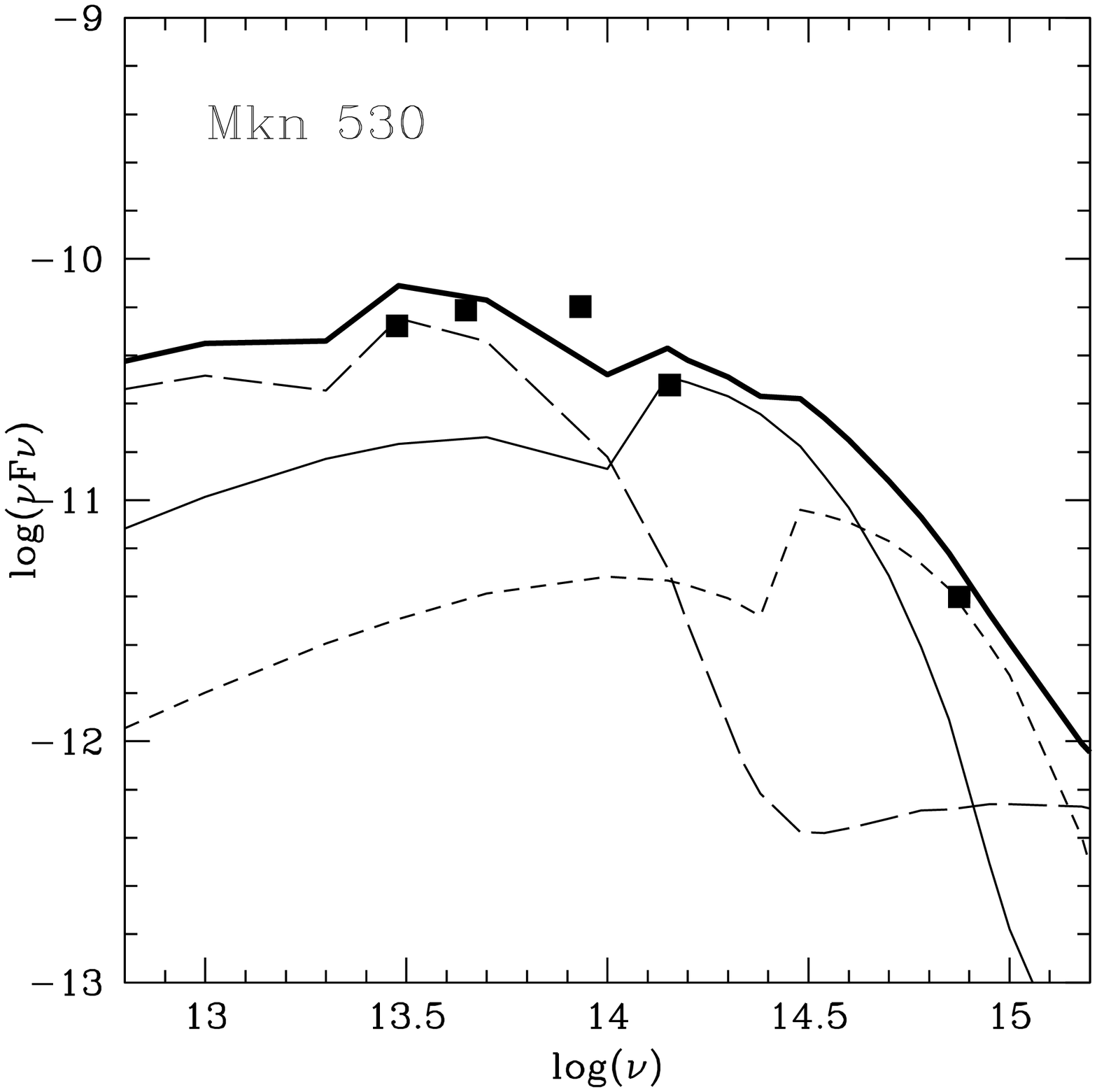}
\caption
{The modelling of selected galaxies from Alonso-Herrero et al. sample:
Seyfert 1-1.5
}
\end{figure*}

\begin{figure*}
\includegraphics[width=78mm]{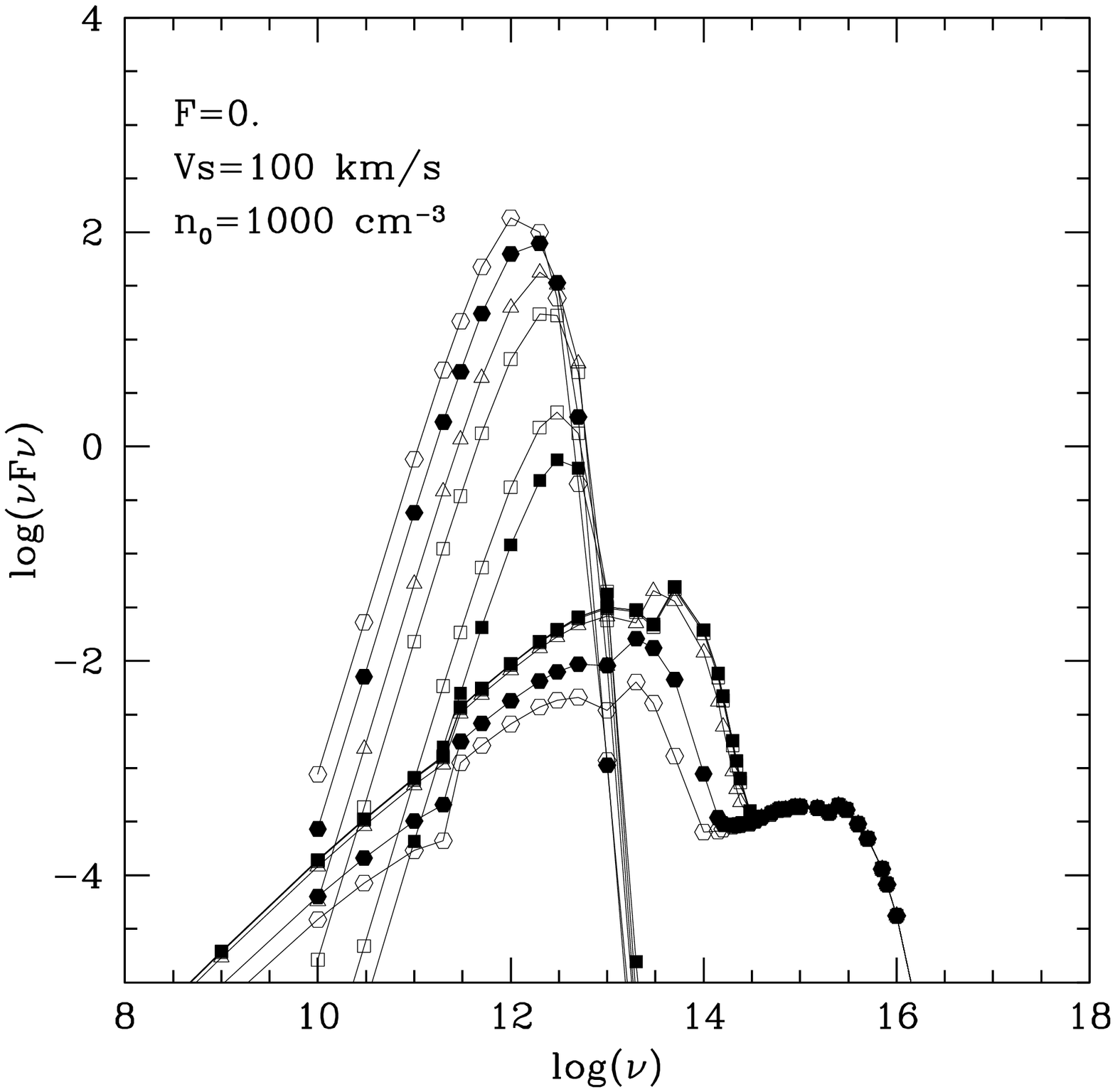}
\includegraphics[width=78mm]{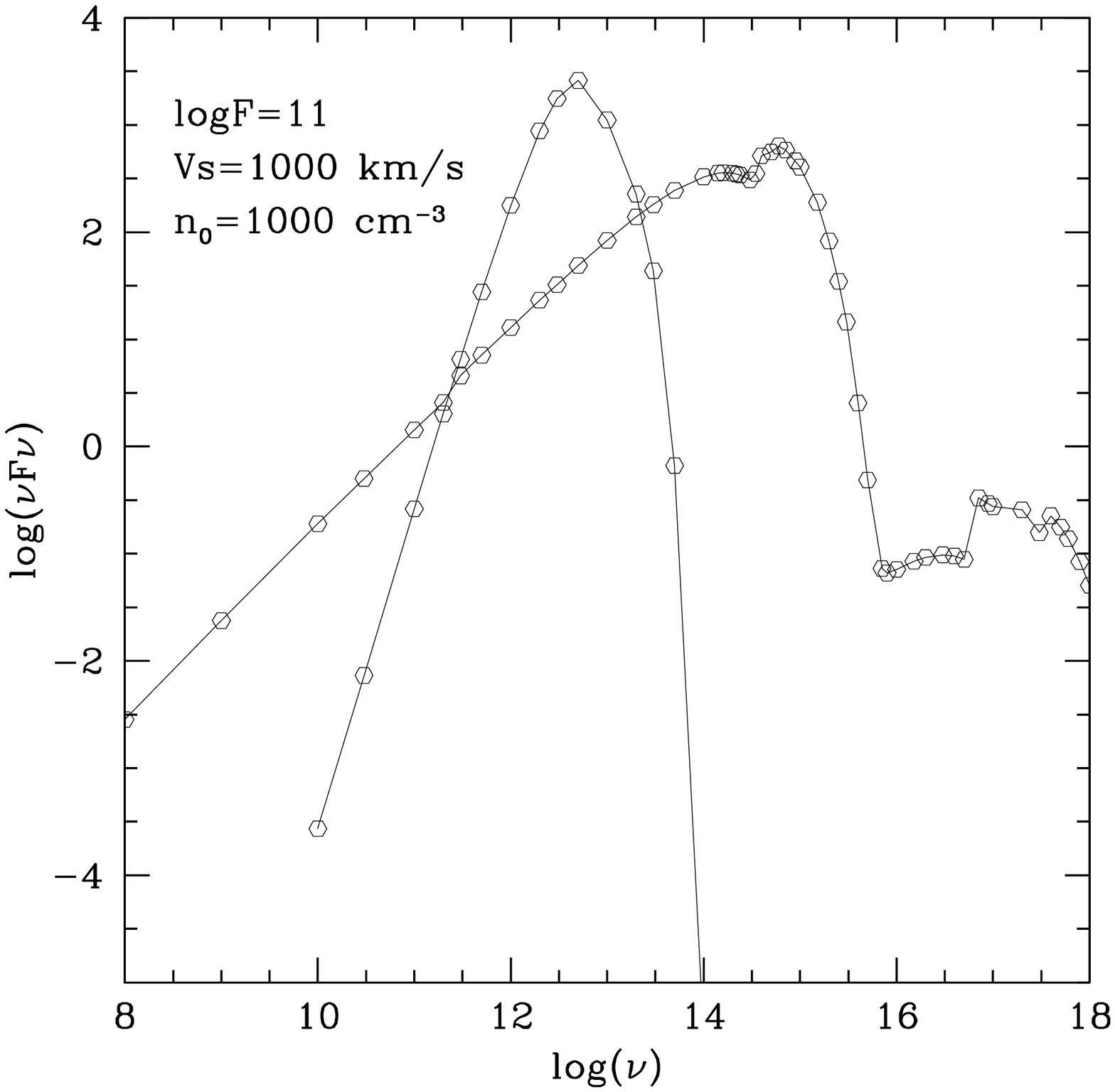}
\caption{The continua obtained from models M1 (left  panel)
and M7 (right panel).  
\agr=0.01 : filled squares, \agr=0.02\mum :
white square, \agr=0.1\mum : filled triangles,
\agr=0.2\mum : white triangles, \agr=0.5\mum : filled circles,
\agr=1.0\mum : white circles.
}

\end{figure*}

\section{Concluding Remarks}

In the last few years we have shown that composite models that account
for the
simultaneous effect of photoionization by a central source and by shocks
produce a fair representation of  the SED  of quite a large number of
different  AGN types, including  Seyfert 1.5-2 galaxies, NLS1, galaxies
where
the AGN coexists with starbursts, active galaxies with very extended
regions and  luminous infrared galaxies.

 In comparison with other models  in the
literature, our approach does not resort to any particular  dust/gas
geometry but explore the possibilities of  two  mechanisms expected
to be present in any AGN,
bremsstrhalung and dust reprocessing, to  reproduce the  IR-optical SED.

We discuss the different contributions to the IR continuum of AGN
considering that both photoionization from the active center and
shocks ionize and heat gas and dust within the surrounding AGN clouds.
Collisional processes start to be important in the presence of shocks
with velocities of  about 100 \kms.  Therefore dust grains are heated by
collision with gas to relatively high temperatures, depending on \Vs.
Moreover, the grains are sputtered crossing the shock front to
different radius depending on the grain initial radius, the plasma
density, and the slab width covered by the grains before being
decelerated and stopped by collisional drag.  The geometrical
thickness of the slabs is determined by the assumption that the
physical conditions in the slab are constant, and are constrained by
the choice of \Vs, \n0, and a$_{gr}$ suitable to the NLR conditions.

Observations indicate that there
is a gradient of densities and velocities in the NLR, both increasing
towards the active center. In the internal region of the NLR the
radiation should be more intense.  We found, however, that in most
Seyfert type 2 objects the high velocity-density clouds are
shock-dominated.  If these clouds are closer to the centre, the effect
of radiation must have been partly cancelled by the effect
of an intervening dusty
medium. Such dusty medium could be clouds with  a very high d/g, which will
emit in the far-infrared.   On the other hand, the  clouds
contributing to intermediate and type 1 Seyfert galaxies   are mainly
radiation-dominated. These clouds have a direct
view of the central radiation as expected from the unified model.

In this paper, the model approach is applied to a sample of galaxies
of different Seyfert types for which the best estimates of the nuclear-
stellar subtracted- optical-IR SED are available.  With a limited range
of parameter values  - shock velocity, pre-shock density, ionizing flux,
grain size and dust-to-gas ratio - which are dictated from our previous
detailed  modelling of representative Seyfert galaxies - the model results
are encouraging as they produce a fair representation of  the SED for all
Seyfert type, in particular,
they easily produce steep IR SED for broad-line type objects.

Further, a first estimate of the dust-to-gas ratio for some of the
clouds is provided. Still, far-IR data  are needed to constrain better
those values.
Fine tunning of the modelling can be
achieved by cross cheking it with the respective
nebular spectrum and the SED at other wavelengths.  Nevertheless,
the analysis of the optical-IR continuum of a  sample of galaxies
provides a general view of the mechanisms producing the emission,
as well as on how they differ in the two types of Seyfert galaxies.

A comparison of Seyfert 1.8-2 with Seyfert 1.5-1 SED models confirms
that continuum fluxes are generally higher for Seyfert 1.5-1 and that
the flux decrease at higher frequencies can be reproduced for most
Seyfert 1.8-2 galaxies.  Recall that in multi-cloud models the
single-cloud models are summed up adopting relative weights.  We
explain, therefore, the flux decrease at higher frequencies by lower
relative weights of clouds with higher velocities and/or reached by a
strong flux from the active center.  

For all type of galaxies, the continuum between 1 and 10 \mum is found to
be  produced by 
 thermal bremsstrahlung emitted by gas with relatively high density,
\n0 ~(1000\cm3), heated by a  low  velocity shock, \Vs ~(100 \kms). These
clouds produce  weak lines and therefore their presence can  mostly be
accessed from the SED analysis.  Because of their low velocity, we expect
that dust  re-emits in the far-infrared.

To provide a first analysis of an AGN SED,
an analytical method is presented in the Appendix.
In general, galaxy continua are explained by multi-cloud
models, leading to extended IR bumps, which must be decomposed
into single components.
Indeed, the SED profile in the IR of a galaxy depends on the  
ratio between the dust
reradiation peak and the bremsstrahlung emission in the IR,
i.e. the dust-to-gas ratios
in the single-cloud models, but depends also on
the relative weights adopted to sum up the different contributions,
which  are constrained by the line spectra.

A linear fit of the bremsstrahlung
emission (normalized to the value at
12 \mum) is adequate for log($\nu$) $<$ 13.50, while the emission by
dust can be represented by a black body (see Appendix).

\section*{Acknowledgments} We are thankful  to an anonymous referee
for the comments which helped to improve this paper.
This research has  been partly supported by the Brazilian
agencies FAPESP(00/06695-0) and CNPq(304077/77-1.

\appendix

\section{Analytical approach}

To illustrate the procedure on how to fit the
SED of a given galaxy using the present modelling, an example  for the
case of the
SED of Arp 220  is given below.
Arp 220 was chosen for two reasons: a rich dataset is
available for this galaxy and a single-cloud model is enough to
provide a good fit of the observed continuum from radio to ultraviolet.

The observational data come from the NED (NASA/IPAC EXTRAGALACTIC DATABASE)
(Benford 1999, Carico et al. 1992,
Chini et al. 1986, Condon et al, 1983, De Voucouleurs et al. 1991,
Douglas et al. 1996,
Dressel \& Condon 1978, Dunne et al. 2000, Eales et al. 1989, Gregory \&
Condon 1991,
Jarrett et al. 2003,  
Moshir et al. 1990, Rigopolous et al. 1996, Soifer et al. 1989,
Spinoglio et al.  1995,
White \& Becker 1992, Waldram et al. 1996, Zwicky \& Herzog 1963)
and from ISOPHOT (Spinoglio et al. 2002)
The SED of Arp 220 (Fig. A1) shows a shape similar to that of model M7
(Table 1, Fig. 7). Bremsstrahlung explains the data
in the radio, optical and soft-X ray  
after removing the  background stellar  contribution.
In the IR, a  single black-body type bump fits the NED (black squares)
data as well as those from ISOPHOT (white squares).
The  SED  can be accounted for with clouds that have
\Vs=1000 \kms,  \n0=1000 \cm3, and a$_{gr}$ = 1 \mum.
The d/g ratio is determined by   shifting the IR bump along the
verical axis
by about  one order relative to  the original model M7.
Note that  model M7 is originally calculated for  d/g=10$^{-13}$,
leading  to a too low IR bump (Fig. A1, thin dotted line).
The IR data are explained, therefore, with d/g $\sim$ 10$^{-12}$,
which is relatively high, in agreement with the fact  that Arp 220 is
 very dusty
(Arp et al. 2001) and is also classified  as an ULIRG.

\begin{figure}
\includegraphics[width=78mm]{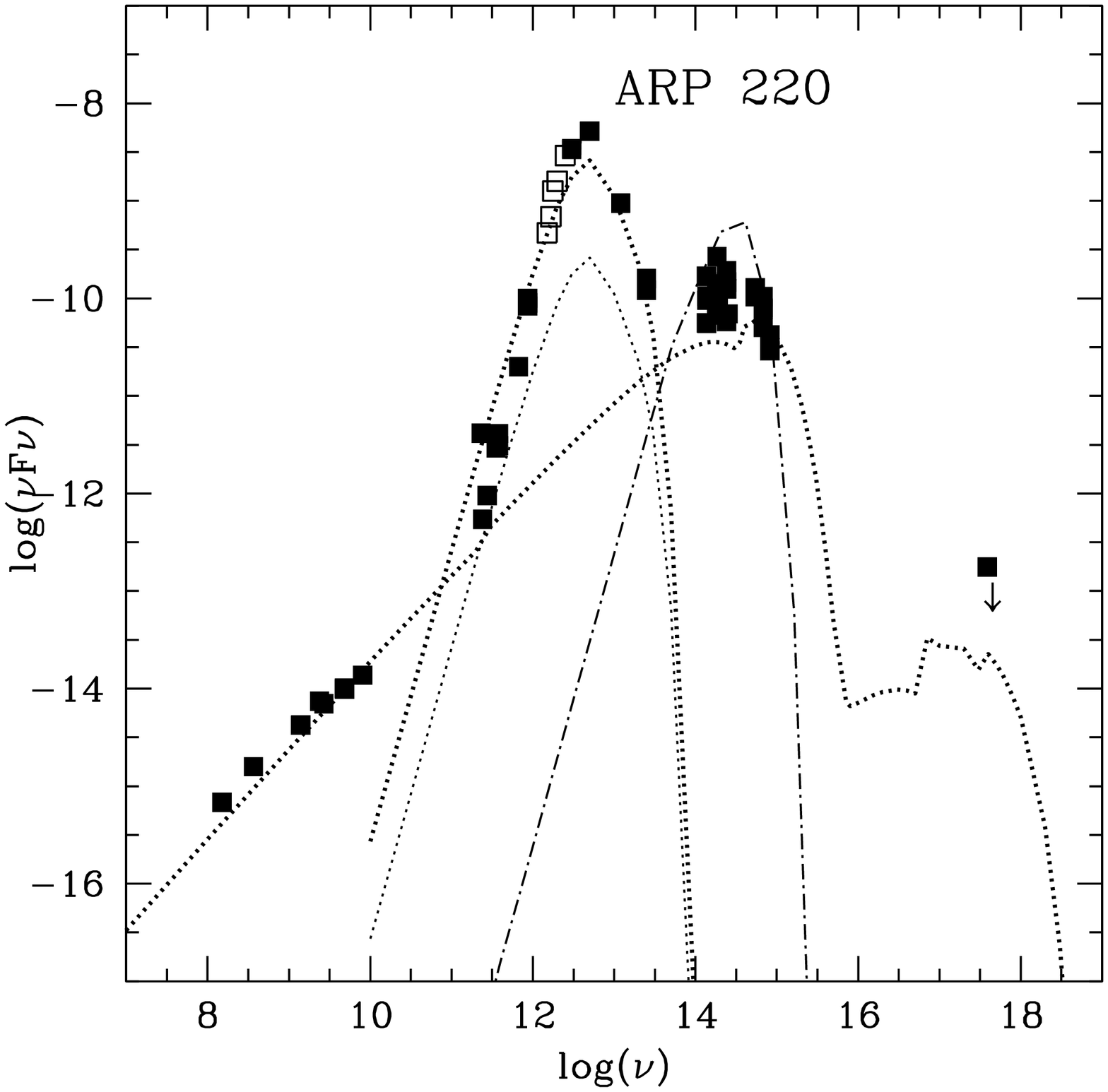}
\caption{The continuum of ARP 220. Symbols as in Fig. 7}
\end{figure}

In order to allow interested readers to use our model approach,
a grid of  single-cloud
models is provided. In all models
it is assumed that the shock acts on the edge opposite to that
reached by the ionizing photons from the nuclear source,  i.e.,
clouds are moving outwards.The following input parameters are kept
constant as follows: the preshock density, \n0 = 300
cm$^{-3}$, the geometrical thickness of the cloud, D = 10$^{19}$ cm,
the shape of the ionizing radiation: a power law with power index
equal to 1.5, the dust-to-gas ratio, d/g = 10$^{-14}$ by number, and
the magnetic field, B$_0$ =10$^{-4}$ gauss.  The elemental abundances
correspond to the cosmic values.  On the other hand, the shock
velocity varies in the range 100 $\leq$ \Vs $\leq$ 900 \kms,
the initial radius of the dust grain in the range 0.01 $\leq$ \agr
$\leq$ 1.0 \mum, and the flux at the Lyman limit \Fh is 10$^{11}$
and 10$^{13}$ cm$^{-2}$ s$^{-1}$ eV$^{-1}$ (Table A1).

The  SED of galaxies,  usually  show the old stellar population
contribution
between about 3 10$^{14}$ and 3 10$^{15}$ Hz.
The data are nested between the black body curve corresponding
to a temperature of T= 3000 K (upper limit) and the
bremsstrahlung (lower limit).
Not always the lower limit is sharply defined.
Nevertheless, the lower limit of the data in this  
range of  frequencies roughly constrains the bremsstrahlung.
It is, therefore,  recommended to use a large range of frequencies,
e.g., 10$^8$-10$^{19}$ Hz
to obtain consistent modelling  in the different
domains (radio, IR, optical, UV, and softX-ray).
In Figs. A2 and A3  we present the  SEDs calculated
for models which appear in Table A1.  
The model SEDs are calculated at the nebula.
In order to compare them
with the data the models must be shifted by a factor which depends on the
distance of the galaxy, the distance of the cloud from the center, and the
covering factor (see Contini, Viegas \& Campos  2003).
Then the bremsstrahlung is adjusted considering the  
value at the maximum frequency, which depends on \Vs
and the maximum in the optical range which depends on \Fh.
The model is constrained by the IR peak frequency,
which also depends on \Vs.

Notice that the models are calculated with d/g = 10$^{-14}$ by number,
so, the shift of the IR bump determins the d/g characteristic of the cloud.
Complex shapes of the bremsstrahlung maximum and of the
IR bump imply a composite model, so the other components
must be chosen from   Figs. A2 e A3  and shifted accordingly.

In the following, based on these models, an analytical method
is proposed as an alternative for using the SEDs given above.
For each model, the free-free and dust emissions are normalized to
their  respective values at 12 \mum.
These values (N$_{br}$ and N$_{dust}$
corresponding to the log($\nu$F$_{\nu}$) at 12 \mum ) are listed in
Table A1.   Each of the models in table A1 has a characteristics \Fh,
\Vs and \agr,
for wich the corresponding black-body effective dust temperature
T$_{dust}$ is given in Table A2.

On the other hand, a
linear fit of the logarithm of the normalized bremsstrahlung emission
is obtained for 11.5 $  < $ log($\nu$) $  < $ log($\nu_{lim}$). In Table A3
the coefficients for the linear fit (A and B) and log($\nu_{lim}$)
are listed for each model.  As seen in Fig. 1, for high frequencies,
the behaviour of the free-free emission depends on the input
parameters of the models and is in general characterized by two maxima and a
minimum value. In Table A4, for each model the frequency and the
$\nu$F$_{\nu}$ is given for those points identified as max$_1$,
min$_1$, and max$_2$ (Fig. 1), as well as the flux at
log $\nu$ = 15.0.

Using the results listed in Tables A1, A2, and A3,  the normalized gas
emission can be obtained between
11.5 $  < $ log($\nu$) $  < $ log($\nu_{lim}$).
The continuum from dust normalised (yd$_{\nu}$) is a black body
with T=T$_{dust}$, i.e.
yd$_{\nu}$ = log($\nu$F$_{\nu}$/$\nu_{12}$F$_{\nu_{12}}$)
with   log($\nu_{12}$)  = 13.40, and  Planck function in photons
cm$^{-2}$ s$^{-1}$:

\begin{table}
\caption{The logarithmic values of $\nu$F$_{\nu}$ at 12 \mum ~for
the bremsstrahlung and dust emission}
\small{
\begin{tabular}{ccccccccc}\\ \hline
F$_h$ & \Vs & a$_{gr}$  &   N$_{br}$   &   N$_{dust}$\\
1e11 &100  & 0.01   &    -0.1330   &    5.2637\\
1e11 &100  & 0.02   &    -0.1330   &   -5.4456\\
1e11 &100  & 0.10   &    -0.1330   &   -5.4569 \\
1e11 &100  & 0.20   &    -0.1330   &   -5.7847\\
1e11 &100  & 0.50   &    -0.1330   &   -9.6800\\
1e11 &100 &  1.00   &    -0.1330   &   - ------\\
&&&&\\
1e11 &300  & 0.01   &    -0.8259   &   -0.2872\\
1e11 &300  & 0.02   &    -0.8279   &   -0.2419\\
1e11 &300  & 0.10   &    -0.6057   &   -0.1440\\
1e11 &300  & 0.20   &    -0.0191   &   -0.2200\\
1e11 &300  & 0.50   &    +1.2140   &   +0.6850\\
1e11 &300  & 1.00   &    -0.9478   &   -2.5944\\
&&&&\\
1e11 &500  & 0.01   &    +1.3327   &   -------\\
1e11 &500  & 0.02   &    -0.3725   &   +0.5220\\
1e11 &500  & 0.10   &    +1.3327   &   +0.8780\\
1e11 &500  & 0.20   &    +0.7107   &   +0.7430\\
1e11 &500  & 0.50   &    +1.2140   &   +0.6960\\
1e11 &500  & 1.00   &    +1.2978   &   +0.9700\\
&&&&\\
1e11 &700  & 0.01   &    +2.2402   &   ------\\
1e11 &700  & 0.02   &    +2.3586   &   ------\\
1e11 &700  & 0.10   &    +2.3586   &   +1.7931\\
1e11 &700  & 0.20   &    +1.8468   &   +1.4480\\
1e11 &700  & 0.50   &    +2.2211   &   +2.0000\\
1e11 &700  & 1.00   &    +2.3220   &   +2.4300\\
&&&&\\
1e11 &900  & 0.01   &    +2.5811   &   ------\\
1e11 &900  & 0.02   &    +2.5811   &   ------\\
1e11 &900  & 0.10   &    +2.4221   &   +2.2097\\
1e11 &900  & 0.20   &    +2.5800   &   +2.1688\\
1e11 &900  & 0.50   &    +2.4221   &   +1.9556\\
1e11 &900  & 1.00   &    +2.5927   &   +2.7200\\
&&&&\\
1e13 &100  & all    &    +0.5297   &   -0.2506\\
&&&&\\
1e13 &300 &  0.01   &    +1.4734   &   +0.2985\\
1e13 &300 &  0.02   &    +1.4807   &   +0.3734\\
1e13 &300 &  0.10   &    +1.4818   &   +0.4438\\
1e13 &300 &  0.20   &    +1.4818   &   +0.3969\\
1e13 &300 &  0.50   &    +0.9388   &   +0.1656\\
1e13 &300 &  1.00   &    +1.1007   &   -2.5953\\
&&&&\\
1e13 &500  & 0.01   &    +1.4732   &   ------\\
1e13 &500  & 0.02   &    +1.4708   &   +0.5727\\
1e13 &500  & 0.10   &    +1.6263   &   +0.6687\\
1e13 &500  & 0.20   &    +1.5026   &   +1.3632\\
1e13 &500  & 0.50   &    +1.5062   &   +1.1831\\
1e13 &500  & 1.00   &    +1.8973   &   +1.5431\\
&&&&\\
1e13 &700  & 0.01   &    +2.3270    &  ------\\
1e13 &700  & 0.02   &    +2.3339    &  ------\\
1e13 &700  & 0.10   &    +2.3307    &  +1.7810\\
1e13 &700  & 0.20   &    +1.8735    &  +1.4395\\
1e13 &700  & 0.50   &    +2.2000    &  +1.9817\\
1e13 &700  & 1.00   &    +2.2968    &  +2.4134\\
&&&&\\
1e13 &900 &  0.01   &    +2.3983   &   ------\\
1e13 &900 &  0.02   &    +2.4116   &   ------\\
1e13 &900 &  0.10   &    +2.3768   &   +1.4790\\
1e13 &900 &  0.20   &    +2.3977   &   +1.8422\\
1e13 &900 &  0.50   &    +2.5062   &   +2.3490\\
1e13 &900 &  1.00   &    +2.5584   &   +2.7273\\
\hline

\end{tabular}}

\end{table}

\begin{table}
\caption{Effective dust temperature}
\begin{tabular}{ccccccc}\\ \hline
\  \Fh=1e11 &&&&&&\\
\  \Vs/a$_{gr}$ &  0.01 &   0.02  &   0.10  &  0.20   &  0.50   & 1.00 \\
\ 100    &       85.  &      70. &     62. &    56.  &    40.  &   36.\\
\ 300    &      230.  &     214. &    214. &   214.  &   234.  &   39.\\
\ 500    &      0.0   &     282. &    270. &   270.  &   225.  &  170.\\
\ 700    &      0.0   &     0.0  &    340. &   324.  &   288.  &  275.
\\                
\ 900    &      0.0   &     0.0  &    340. &   324.  &   324   &  300.\\
\  \Fh=1e13 &&&&&&\\
\ \Vs/a$_{gr}$  &  0.01 &   0.02  &   0.10   & 0.20   &  0.50  &  1.00 \\
\ 100     &      170.   &    170. &    170. &   148.  &   78.  &   18.\\
\ 300     &      240.   &    224. &    186.  &  230.  &  182.  &   42.  \\
\ 500     &      0.0    &    316. &    295.  &  270.  &  270.  &  257.\\
\ 700     &      0.0    &    0.0  &    309.  &  323.  &  288.  &  270.\\
\ 900     &      0.0    &    0.0  &    340.  &  330.  &  310.  &  310.\\
\hline
\end{tabular}
\end{table}

\begin{table}
\caption{Linear fit to the logarithm of the
normalized bremsstrahlung emission:  y = A log($\nu$) + B }
\begin{tabular}{ccccc}
\hline
\    \Fh  &    \Vs    & log($\nu_{lim}$)&    A    &  B    \\
\ 1e11  &   100    &     13.30    &      0.77  &  -10.42\\
\ 1e11  &   300    &     13.10    &      0.80  &  -10.61\\
\ 1e11  &   500    &     13.10    &      0.80  &  -10.60\\
\ 1e11  &   700    &     13.30    &      0.75  &  -10.05\\
\ 1e11  &   900    &     13.30    &      0.76  &  -10.15\\
&&&&\\
\ 1e13   &  100    &     13.40    &      0.82  &  -10.98\\
\ 1e13   &  300    &     13.40    &      0.76  &  -10.15\\
\ 1e13   &  500    &     13.40    &      0.76  &  -10.14\\
\ 1e13   &  700    &     13.30    &      0.81  &  -10.76\\
\ 1e13   &  900    &     13.30    &      0.77  &  -10.24\\
\hline
\end{tabular}
\end{table}

\newpage

\begin{table*}
\caption{Characteristics of the bremsstrahlung emission for log($\nu)
 >$ 13.50}
\begin{tabular}{ccccccc} \hline

\  F$_h$ & \Vs & a$_{gr}$ &  max$_1$ &    min$_1$  &    max$_2$    & last \\
\hline
\ 1e11 &100 &  all  & 14.00/+0.033 &14.28/-0.142 &14.36/+0.344
&15.00/-0.654  \\
&&&&&&\\
\ 1e11 &300 &  0.01 & 13.90/+0.123 &14.20/+0.079 &14.36/+0.382
&15.00/-0.664  \\
\ 1e11 &300  & 0.02 & 13.90/+0.123 &14.20/+0.079 &14.36/+0.382
&15.00/-0.664  \\
\ 1e11 &300  & 0.10 & 13.90/+0.123 &14.20/+0.079 &14.36/+0.382
&15.00/-0.664  \\
\ 1e11 &300  & 0.20 & 14.00/+0.152 &14.25/+0.620 &14.36/+0.460
&15.00/-0.581\\
\ 1e11 &300  & 0.50 & 14.00/+0.230 &14.37/+0.096 &14.48/+0.515
&15.00/-0.141\\
\ 1e11 &300  & 1.00 & 14.15/+0.230 &14.30/+0.118 &14.36/+0.475
&15.00/-0.568\\
\ &&&&&&\\
\ 1e11 &500  & 0.01 & 13.70/+0.075 &14.00/-0.003 &14.14/+0.316
&15.00/-1.000\\
\ 1e11 &500  & 0.02 & 13.70/+0.075 &14.00/-0.003 &14.14/+0.316
&15.00/-1.000  \\
\ 1e11 &500  & 0.10 & 13.70/+0.075 &14.00/-0.003 &14.14/+0.316
&15.00/-1.000\\
\ 1e11 &500  & 0.20 & 14.00/+0.174 &14.30/+0.066 &14.36/+0.468
&15.00/-0.469\\
\ 1e11 &500  & 0.50 & 14.00/+0.226 &14.38/+0.092 &14.48/+0.511
&15.00/-0.146\\
\ 1e11 &500  & 1.00 & 14.00/+0.256 &14.38/+0.170 &14.48/+0.558
&15.00/+0.044\\
\ &&&&&&\\
\ 1e11 &700 &  0.01 & 14.10/+0.200 &14.20/+0.146 &14.30/+0.394
&15.00/-0.235\\
\ 1e11 &700  & 0.02 & 14.10/+0.200 &14.20/+0.146 &14.30/+0.394
&15.00/-0.235 \\
\ 1e11 &700  & 0.10 & 14.10/+0.200 &14.20/+0.146 &14.30/+0.394
&15.00/-0.235\\
\ 1e11 &700  & 0.20 & 14.00/+0.186 &14.30/+0.015 &14.36/+0.476
&15.00/-0.500  \\
\ 1e11 &700  & 0.50 & 14.00/+0.186 &14.30/+0.015 &14.36/+0.476
&15.00/-0.500\\  
\ 1e11 &700  & 1.00 & 14.00/+0.217 &14.34/+0.062 &14.48/+0.472
&15.00/-0.270\\
\ &&&&&&\\
\ 1e11& 900 &  0.01 & 14.00/+0.165 &14.20/+0.092 &14.32/+0.433
&15.00/-0.236\\
\ 1e11& 900 &  0.02 & 14.00/+0.165 &14.20/+0.092 &14.32/+0.433
&15.00/-0.236\\
\ 1e11& 900 &  0.10 & 13.76/+0.058 &14.00/-0.098 &14.16/+0.295
&15.00/-1.800  \\
\ 1e11& 900 &  0.20 & 14.99/+0.200 &14.32/+0.084 &14.40/+0.490
&15.00/-0.314\\
\ 1e11& 900 &  0.50 & 14.99/+0.200 &14.32/+0.084 &14.40/+0.490
&15.00/-0.314  \\
\ 1e11& 900 &  1.00 & 14.99/+0.200 &14.32/+0.084 &14.40/+0.490
&15.00/-0.314 \\
\ &&&&&&\\
\ 1e13 &100  & all & 14.36/+0.471 &14.70/+0.412 &14.85/+0.754
&15.00/+0.698 \\
\ &&&&&&\\
\ 1e13 &300  & 0.01  &13.86/+0.128 &14.18/+0.058 &14.30/+0.400
&15.00/-0.504\\
\ 1e13 &300  & 0.02  &13.86/+0.128 &14.18/+0.058 &14.30/+0.400
&15.00/-0.504  \\
\ 1e13 &300  & 0.10  &13.86/+0.128 &14.18/+0.058 &14.30/+0.400
&15.00/-0.504\\
\ 1e13 &300  & 0.20  &13.86/+0.128 &14.18/+0.058 &14.30/+0.400
&15.00/-0.504  \\
\ 1e13 &300  & 0.50  &14.00/+0.235 &14.33/+0.204 &14.53/+0.512
&15.00/+0.066\\
\ 1e13 &300  & 1.00  &13.70/+0.020 &13.85/-0.023 &14.06/+0.410
&15.00/-0.602  \\
\ &&&&&&\\
\ 1e13 &500  & 0.01 & 13.76/+0.117 &14.00/+0.075 &14.33/+0.269
&15.00/-0.560\\
\ 1e13 &500  & 0.02 & 13.76/+0.117 &14.00/+0.075 &14.33/+0.269
&15.00/-0.560  \\
\ 1e13 &500  & 0.10 & 14.36/+0.117 &14.54/+0.235 &14.60/+0.537
&15.00/+0.174\\
\ 1e13 &500  & 0.20 & 13.76/+0.117 &14.00/+0.075 &14.38/+0.320
&15.00/-0.469\\
\ 1e13 &500  & 0.50 & 13.76/+0.117 &14.00/+0.075 &14.38/+0.320
&15.00/-0.469\\
\ 1e13 &500  & 1.00 & 14.00/+0.252 &14.38/+0.100 &14.48/+0.511
&15.00/-0.081 \\
\ &&&&&&\\
\ 1e13 &700 &  0.01 & 14.00/+0.174 &14.15/+0.135 &14.30/+0.412
&15.00/-0.310\\
\ 1e13 &700 &  0.02 & 14.00/+0.174 &14.15/+0.135 &14.30/+0.412
&15.00/-0.310  \\
\ 1e13 &700 &  0.10 & 14.00/+0.174 &14.15/+0.135 &14.30/+0.412
&15.00/-0.310\\
\ 1e13 &700 &  0.20 & 14.00/+0.191 &14.30/+0.040 &14.36/+0.459
&15.00/-0.409 \\
\ 1e13 &700 &  0.50 & 14.00/+0.191 &14.30/+0.040 &14.36/+0.459
&15.00/-0.409   \\  
\ 1e13 &700 &  1.00 & 14.00/+0.221 &14.34/+0.075 &14.48/+0.468
&15.00/-0.228  \\
\ &&&&&&\\
\ 1e13 &900 &  0.01 & 13.70/+0.056 &14.00/-0.508 &14.16/+0.282
&15.00/-1.400\\
\ 1e13 &900 &  0.02 & 13.70/+0.056 &14.00/-0.508 &14.16/+0.282
&15.00/-1.400 \\
\ 1e13 &900 &  0.10 & 13.70/+0.056 &14.00/-0.508 &14.16/+0.282
&15.00/-1.400\\
\ 1e13 &900 &  0.20 & 13.70/+0.056 &14.00/-0.508 &14.16/+0.282
&15.00/-1.400 \\
\ 1e13 &900 &  0.50 & 13.70/+0.100 &14.00/+0.044 &14.16/+0.350
&15.00/-1.000\\
\ 1e13 &900 &  1.00 & 14.00/+0.182 &14.20/+0.122 &14.30/+0.394
&15.00/-0.322\\
\hline\\
\end{tabular}
\end{table*}

\begin{figure*}
\includegraphics[width=78mm]{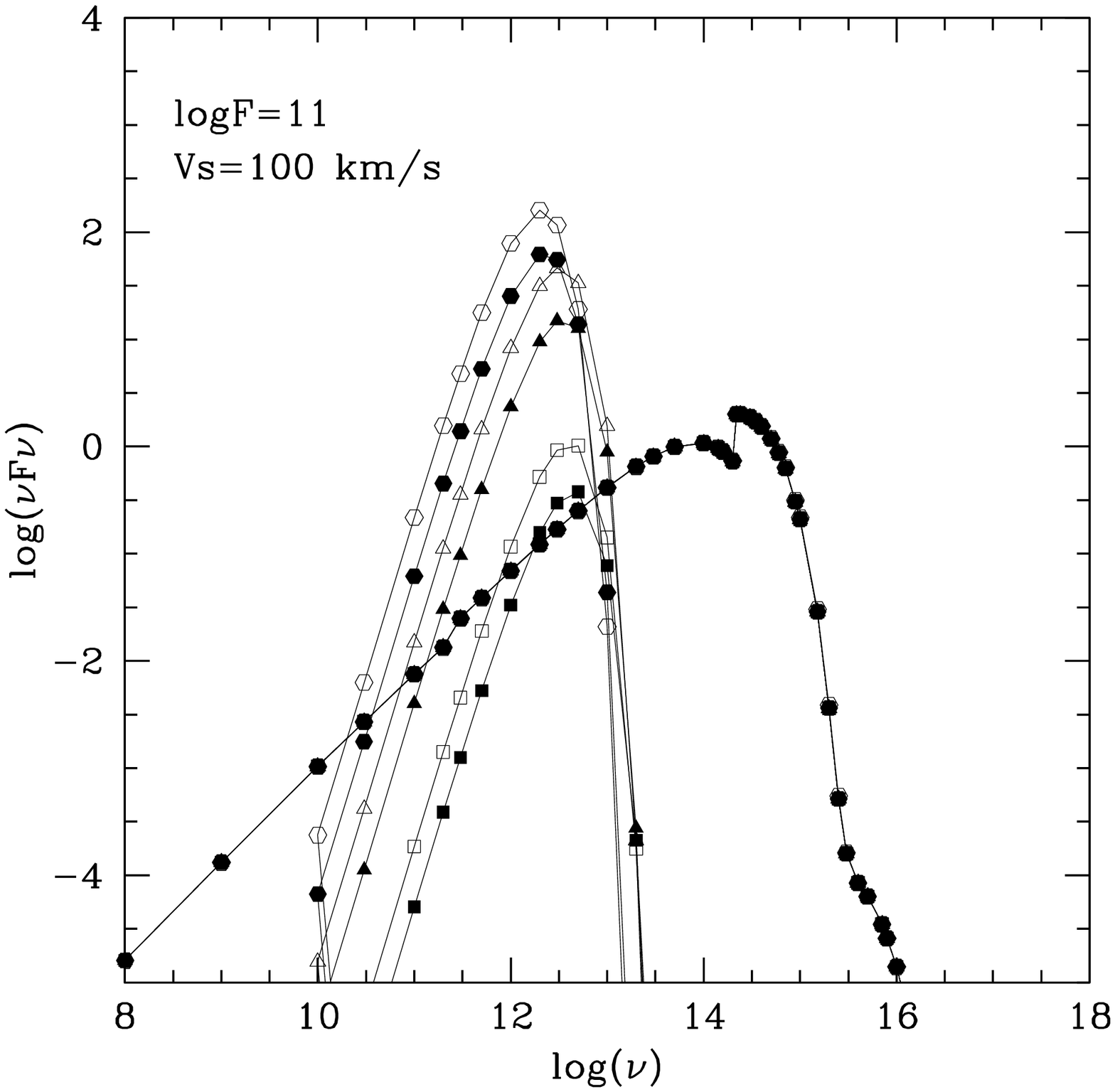}
\includegraphics[width=78mm]{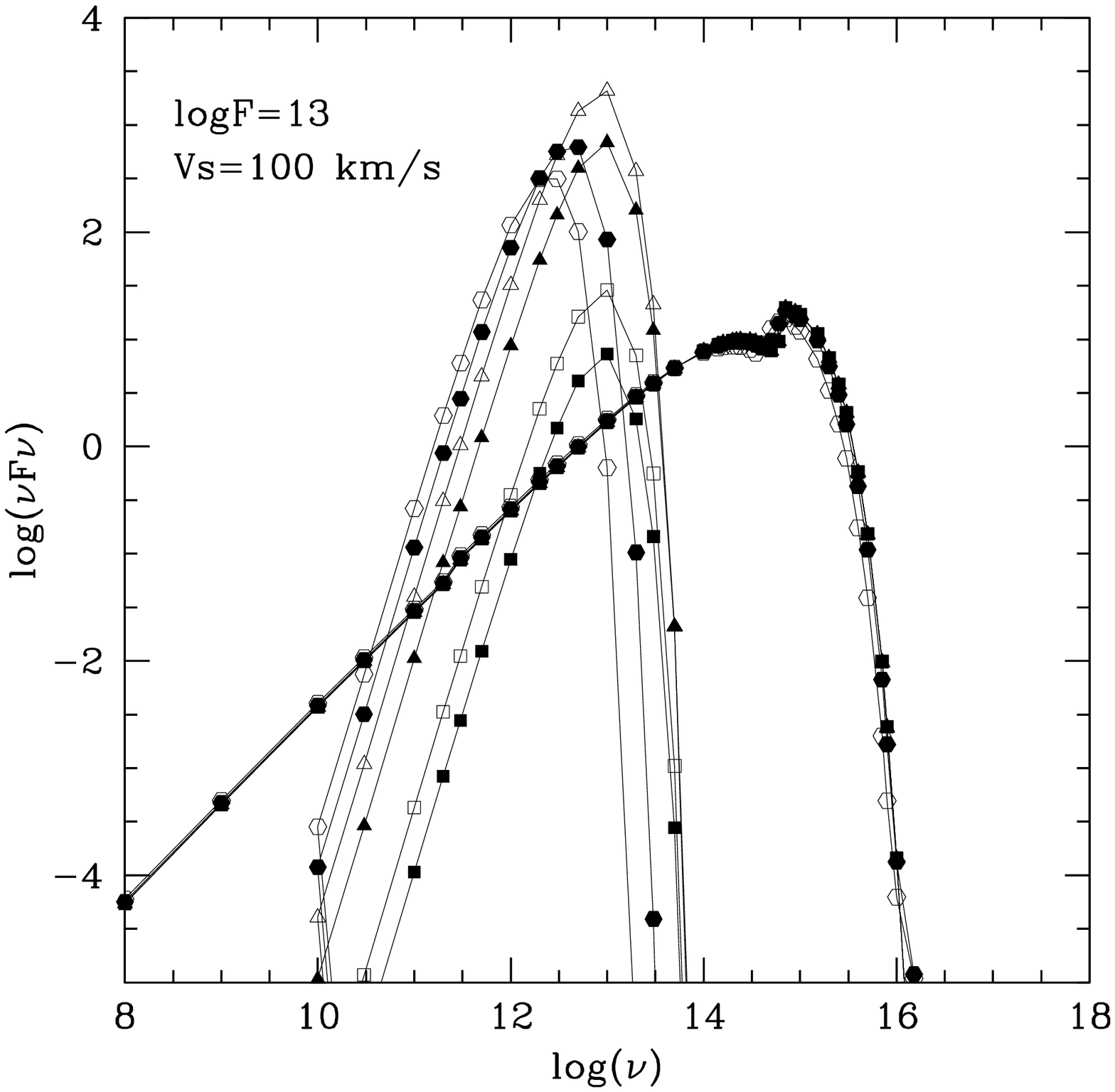}
\includegraphics[width=78mm]{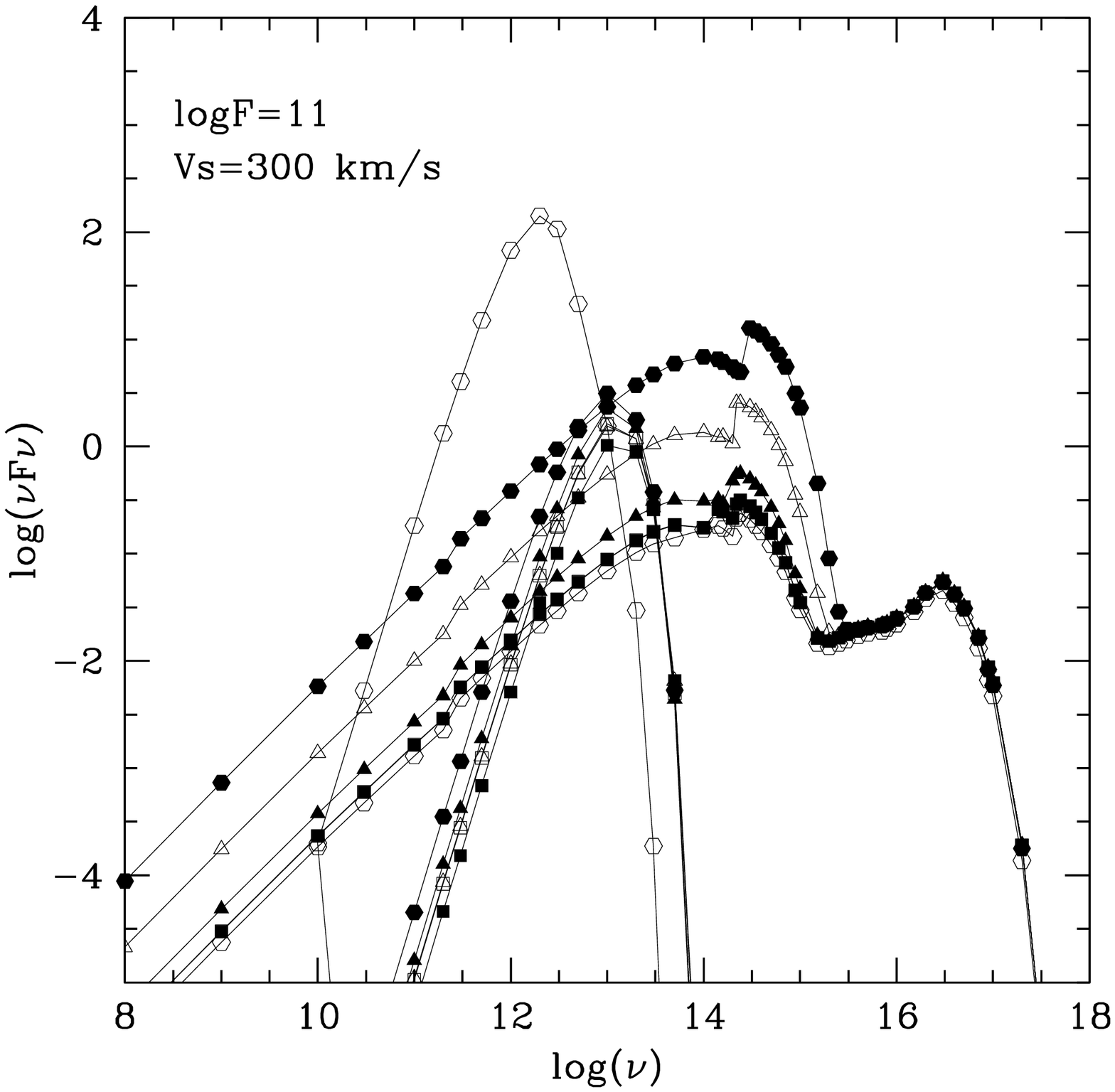}
\includegraphics[width=78mm]{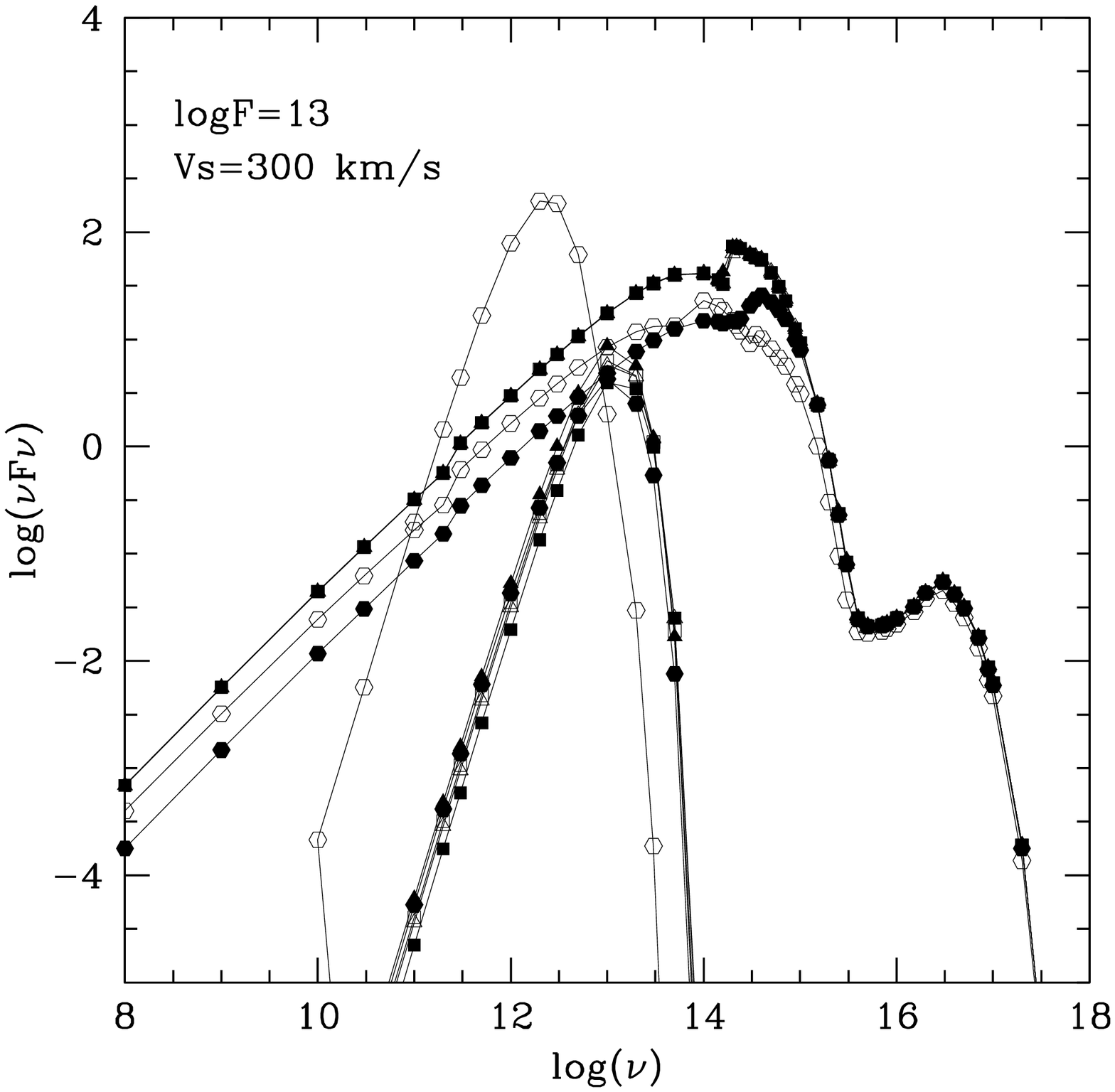}
\includegraphics[width=78mm]{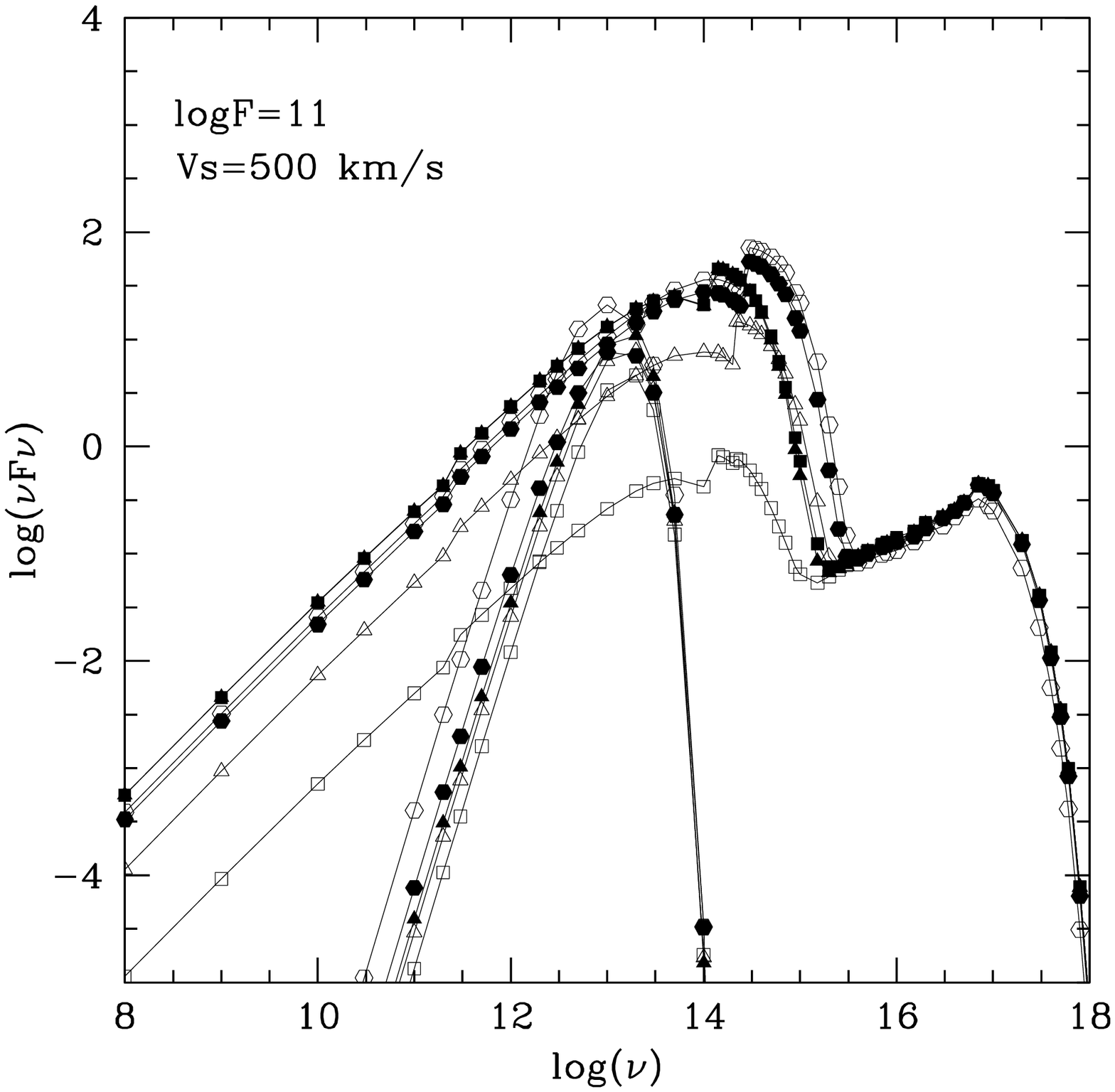}
\includegraphics[width=78mm]{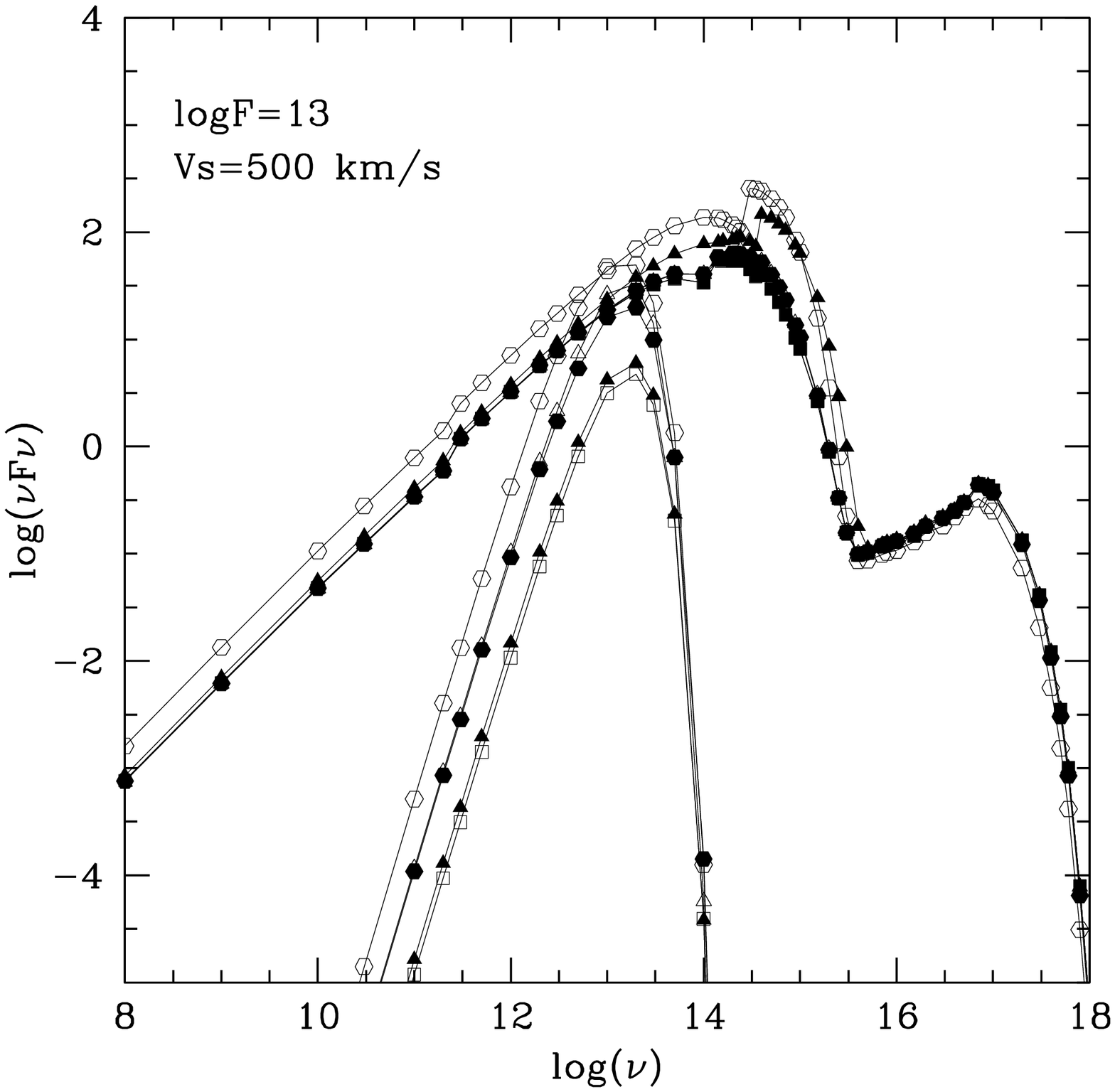}
\caption{Models.Symols as in Fig. 7}
\end{figure*}
\begin{figure*}
\includegraphics[width=78mm]{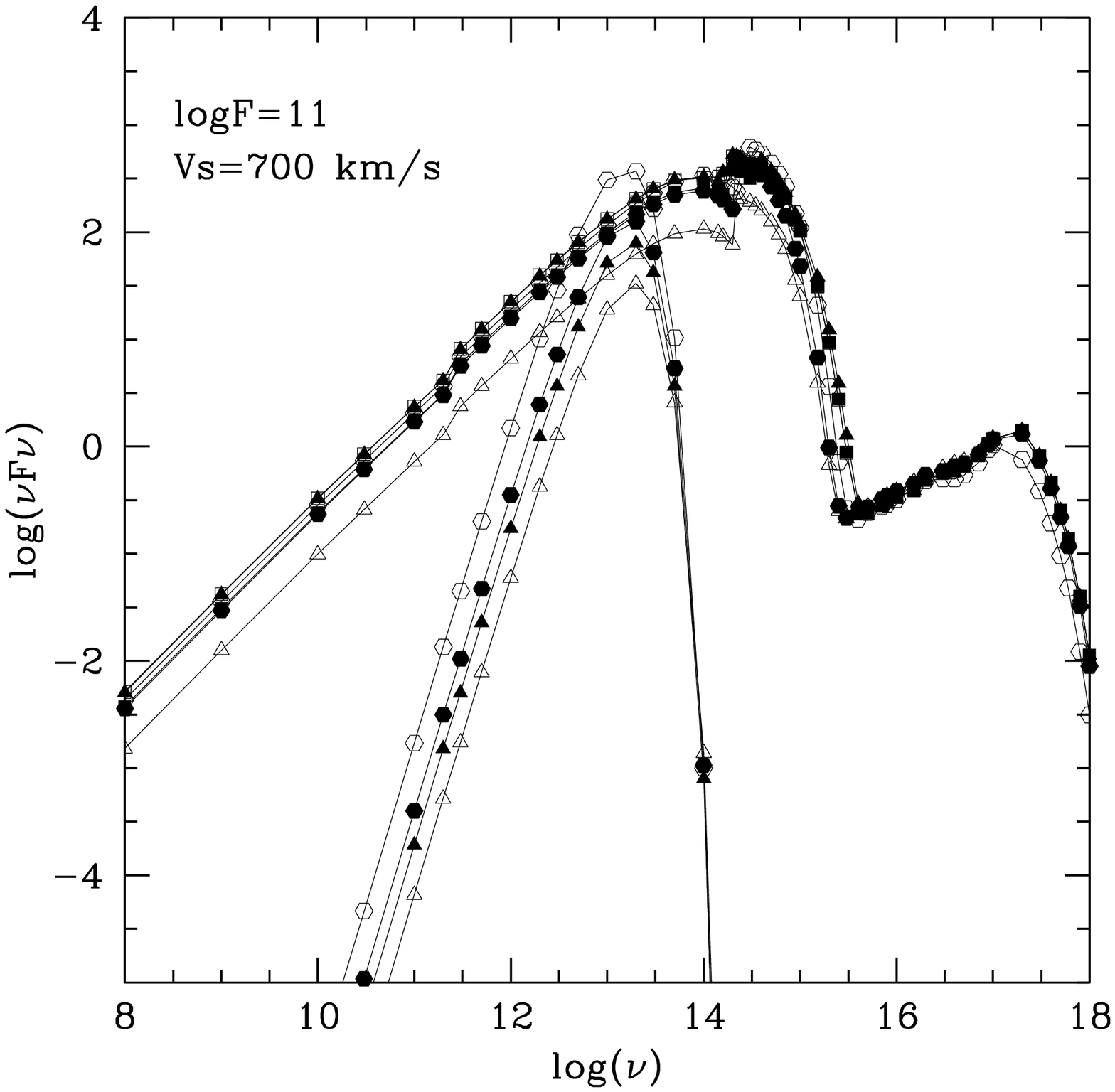}
\includegraphics[width=78mm]{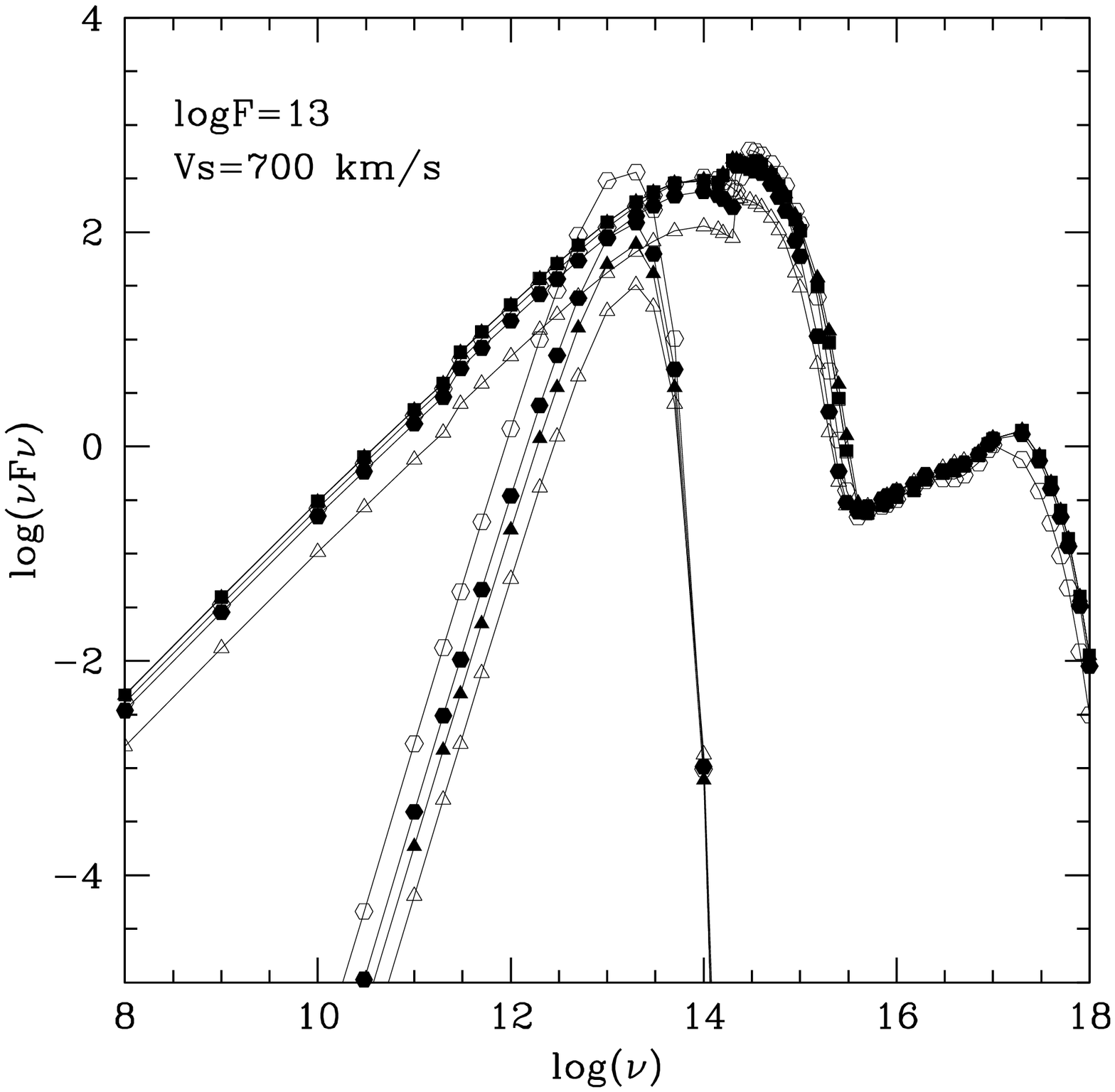}
\includegraphics[width=78mm]{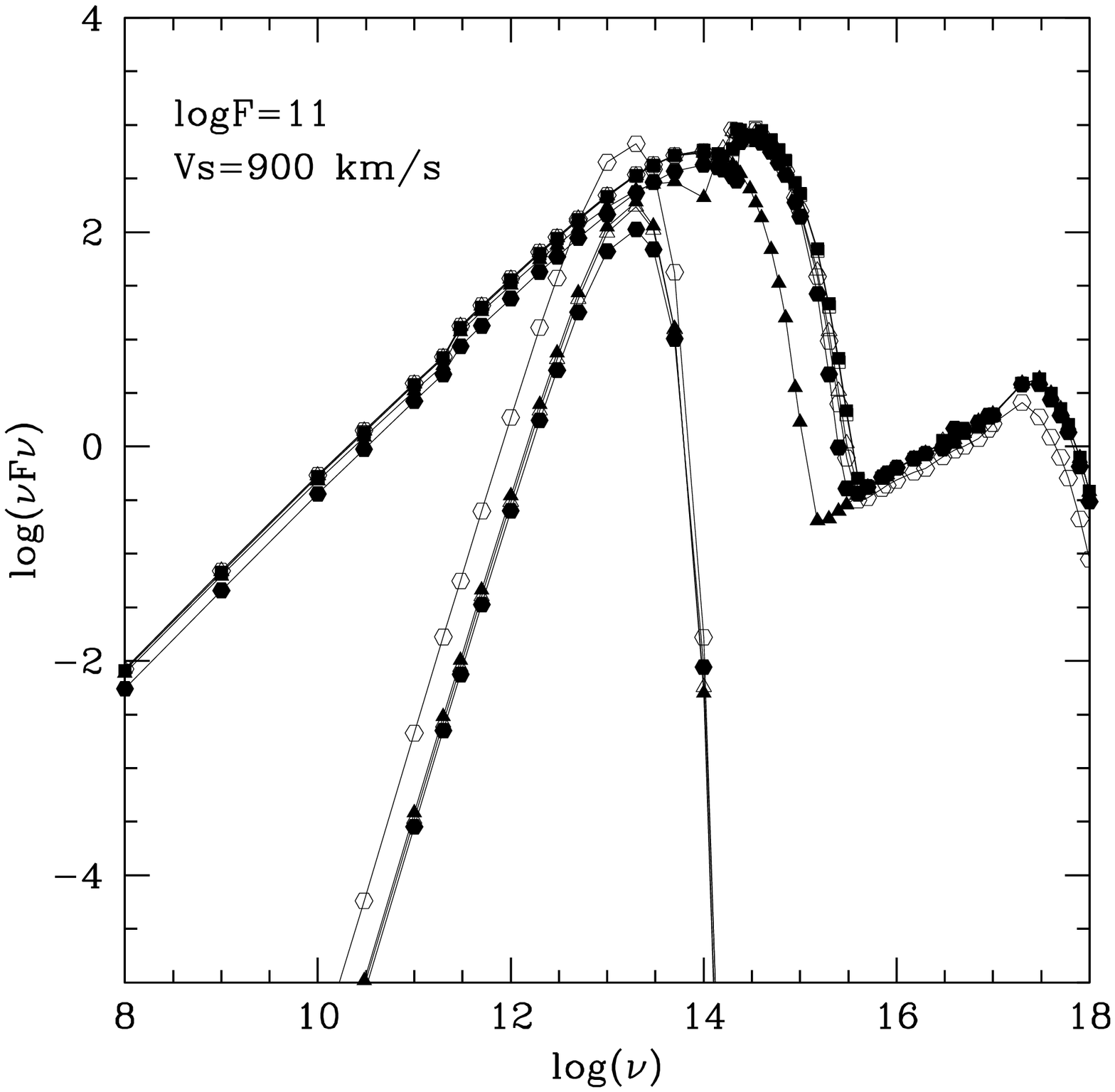}
\includegraphics[width=78mm]{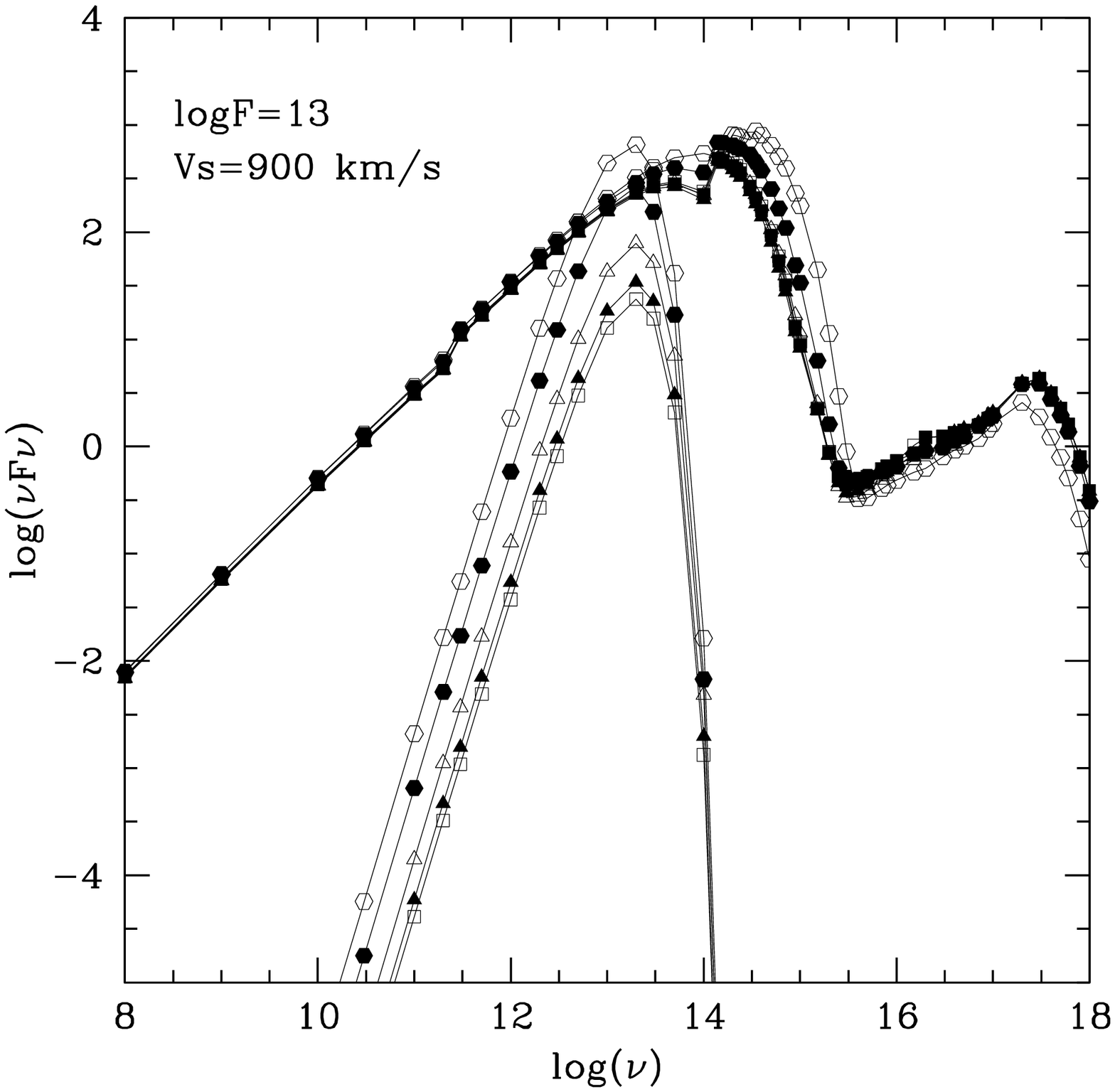}
\caption{Models. Symbols as in Fig. 7}
\end{figure*}

yd$_{\nu}$=3log($\nu/\nu_{12}$)+log((exp(h$\nu_{12}$/(kT$_{dust}$)-1)
/(exp(h$\nu/(kT_{dust}$) -1)),

where   h/k = 4.80 10$^{-11}$. The thermal bremsstrahlung is
obtained from the linear fit.

Notice that for \Vs = 100 \kms the bremsstrahlung emission
is completely independent of \agr.
Nevertheless, the effective dust emission varies with
\agr for all the models. The effective dust temperature depends
on how efficient is the sputtering.
In all
models  the  dust-to-gas ratio is kept constant to  
d/g = 10$^{-14}$ . However, in the fitting procedure of a given galaxy,
d/g shall be modified to account for the observed IR peaks in the mid- and
far-IR SED, as described below.
Note that the importance of
dust relative to free-free emissions depends on the
d/g parameter, particularly in the near-infrared range.

As shown in the previous sections, the infrared-optical SED is
usually due to two or more types of clouds, each one corresponding to
characteristic velocity and grain size. In the following, the steps
that should be followed  to get a crude fit  of the
IR SED,  are given:

(a) the continuum above log($\nu$) $\geq$ 13.50,
is used to find the type of  cloud (\Vs ~and \agr)
explaining the observed data.
The free-free emission dominates the continuum
in that frequency range. Using Tables A3 and A4, the best model is chosen.

(b) once the cloud characteristics  are found, T$_{dust}$, defining the
corresponding black-body curve, is obtained from  Table A2.

(c) because the two curves are normalized by the respective values
at 12 \mum, the non-normalized curves must be used to verify the
relative importance of dust and free-free, by shifting vertically
the black-body component, which corresponds to finding the
d/g for this particular object. Notice that at this point
the scale of the vertical axis is defined by the model,
and only the shape of the curves are used.

(d) The intermediate and far-infrared emission are then
used to obtain other possible  dust components (T$_{dust}$),
which will define the types of cloud (\Vs, \agr)
as well as their relative importance (d/g) by
shifting the non-normalized curves vertically. This will also
define the corresponding free-free components.

(e) The final step is to determine the relative weights
corresponding to each type of cloud to get
the best fit to  the observed flux values.


\begin{thebibliography}{99}

\bibitem{} Allen, C.W. 1973, Astrophysical Quantities (London: Athlon)

\bibitem{} Alonso-Herrero, A., Quillen, A. C., Rieke, G. H., Ivanov, V. D.,
Efstathiou, A., 2003, astro-ph/0303617

\bibitem{} Antonucci, R.R.J. Miller, J.S.  1985, ApJ, 297, 621

\bibitem{} Arp, H.C., Burbridge, E.M., Chu, Y., Zhu, X.  2001, ApJ, 553, L11

\bibitem{} Barvainis,  R., 1992, ApJ, 400, 502

\bibitem{} Benford, D.J. 1999 CIT

\bibitem{} Blain, A.W.,   Barnard, V.E.,   Chapman, S.C.    2003, MNRAS,
338, 733

\bibitem{} Carico, D.P., Keene, J., Soifer, B.T., Neugebauer, G. 1992,
PASP, 104, 1086

\bibitem{} Ciroi, S., Contini, M., Rafanelli, P., Richter, G.M. 2003,
A\&A, 400, 859

\bibitem{} Chini, R., Kreysa, E., Krugel, E., Mezger, P.G. 1986, A\&A,
166, L8

\bibitem{} Condon, J.J., Condon, M.A., Broderick, J.J., Davis, M.M.
1983, AJ, 88, 20

\bibitem{} Contini, M.,  Contini, T. 2003, MNRAS, 342, 299

\bibitem{} Contini, M., Shaviv, G. 1982, ApSS, 85, 203

\bibitem{} Contini, M., Viegas, S. M., 1999  ApJ, ApJ, 523, 114

\bibitem{} Contini, M., Viegas, S. M., 2000, ApJ, 535, 721

\bibitem{} Contini, M., Viegas, S. M., 2001, ApJS, 132, 211

\bibitem{} Contini, M., Viegas-Aldrovandi, S. M., 1990, ApJ, 350, 125

\bibitem{} Contini, M., Prieto, M.A., Viegas, S.M. 1998a, ApJ, 492, 511

\bibitem{} Contini, M., Prieto, M.A., Viegas, S.M. 1998b, ApJ, 505, 621

\bibitem{} Contini, M., Rodriguez-Ardila, A., Viegas, S.M. 2003, A\&A,
in press

\bibitem{} Contini, M., Viegas, S.M., Campos, P.E. 2003 MNRAS, in press

\bibitem{} Contini, M., Viegas, S.M., Prieto M.A. 2002, A\&A, 386, 399

\bibitem{} De Voucouleurs, G., De Voucouleurs, A., Corwin Jr., H.G.,
Buta, R.J., Paturel, G., Fouque, P. 1991, %%@
IRC3.9

\bibitem{} Douglas, J.N., Bash, F.N., Bozyan, A., Torrence, G.W., Wolfe,
C. 1996, AJ, 111, 1945

\bibitem{} Draine, B.T. 1981, ApJ, 245, 880

\bibitem{} Draine, B.T.,  Salpeter, E.E. 1979, ApJ, 231, 438

\bibitem{}  Dressel, L.L., Condon, J.J. 1978, ApJS, 36, 53

\bibitem{} Dunne, L., Eales, S., Edmunds, M., Ivison, R, Alexander, P.,
Clements, D.L. 2000 MNRAS, 315, 115

\bibitem{} Dwek, E. 1981, ApJ, 247, 614

\bibitem{} Dwek, E., Foster, S.M., Vankura, O.  1996, ApJ, 457, 244

\bibitem{} Eales, S.A., Wynn-Williams, C.G., Duncan, W.D. 1989 ApJ, 339, 859

\bibitem{} Edelson, R. A., Malkan, M. A.,  1986, ApJ 308, 59

\bibitem{} Granato, G. L., Danese, L., 1994, MNRAS, 268, 235

\bibitem{} Granato, G. L., Danese, L., Franceschini, A., 1997, ApJ, 486, 147

\bibitem{} Gregory, P.C., Condon, J.J. 1991, ApJS, 75, 1011

\bibitem{} Jarrett, T.H., Chester, T., Cutri, R., Schneider, S.E.,
Huchra, J.P. 2003, AJ, 125, 525

\bibitem{} Laor, A., Draine, B. T., 1993, ApJ, 402, 441

\bibitem{} Moshir, M. et al. 1990, IRASF

\bibitem{} Nenkova, M., Ivezi\'c, Z., Elitzur, M., 2002, ApJ, 570, L9

\bibitem{} Pier, E. A., Krolik, J. H., 1992, ApJ, 401, 99

\bibitem{} Rieke, G.H., Low, F.J. 1975, 199. L13

\bibitem{} Rigopolous, D., Lawrence, A., Rowan-Robinson, A. 1996, MNRAS
278, 1049

\bibitem{} Rowan-Robinson, M.  1992 MNRAS, 258, 787

\bibitem{} Shull, J.M. 1978, ApJ, 226, 858

\bibitem{} Soifer, B.T., Boehmer, L., Neugebauer, G., Sanders, D.B.
1989, AJ, 98, 766

\bibitem{} Spinoglio, L., Andreani, P. Malkan, M.A.   2002, 572, 105

\bibitem{} Spinoglio, L., Malkan, M.A., Rush, B., Carrasco, L.,
Recillas-Cruz, E. 1995 ApJ, 453, 616

\bibitem{} Two micron all sky survey team 2003, 2MASX

\bibitem{} Vaceli, M. S., Viegas, S. M., Gruenwald, R.,
Benevides-Soares, P.,
1993, PASP, 105, 875

\bibitem{} Viegas, S. M., Contini, M., 1994, ApJ, 428, 113

\bibitem{} Viegas, S. M., Contini, M., 1997,  IAU Colloquium
  159 on Emission Lines in Active Galaxies: New Methods and Techniques,
  Eds. B. M. Peterson, F.-Z. Cheng \& A. S. Wilson (San Francisco: A.S.P.)

\bibitem{} White, R.L., Becker, R.H. 1992, ApJS, 79, 331

\bibitem{} Waldram, E.M., Yates, J.A., Riley, J.M., Warner, P.J. 1996
MNRAS, 282, 779

\bibitem{} Zwicky, F., Herzog, E. 1963, CGCG2.

\end{thebibliography}
\end{document}